\documentclass[smallextended]{svjour3}       
\smartqed  
\usepackage{graphicx}

\usepackage{natbib}
\usepackage{url}
\usepackage[english]{babel}
\usepackage{amsmath}
\usepackage{amssymb}
\usepackage{graphicx, subcaption}
\usepackage[colorinlistoftodos]{todonotes}
\usepackage{booktabs,dcolumn,
caption
}
\usepackage{rotating}
\usepackage{moreverb}
\usepackage{bm}
\usepackage{multirow}
\usepackage{tablefootnote}
\usepackage[para]{threeparttable}
\usepackage{csquotes}
\usepackage{float}
\usepackage{enumitem}
\setitemize{leftmargin=0.7in}

\newcolumntype{d}[1]{D{.}{.}{#1}}

\def\bG{{\mathbf G}}
\def\bX{{\mathbf X}}
\def\bZ{{\mathbf Z}}
\def\bM{{\mathbf M}}
\def\bN{{\mathbf N}}
\def\bS{{\mathbf S}}
\def\bB{{\mathbf B}}
\def\bA{{\mathbf A}}
\def\ba{{\mathbf a}}

\def\bH{{\mathbf H}}
\def\bJ{{\mathbf J}}
\def\bW{{\mathbf W}}

\def\bK{{\mathbf K}}
\def\bI{{\mathbf I}}

\def\bE{{\mathbf E}}
\def\bQ{{\mathbf Q}}
\def\bU{{\mathbf U}}
\def\bV{{\mathbf V}}

\def\bde{{\mathbf e}}
\def\bltre{{\mathbf e}}

\def\bse{\begin{eqnarray*}}
\def\ese{\end{eqnarray*}}
\def\be{\begin{eqnarray}}
\def\ee{\end{eqnarray}}
\def\bq{\begin{equation}}
\def\eq{\end{equation}}

\def\bLambda{{\boldsymbol{\Lambda}}}

\def\bbeta{{\boldsymbol{\beta}}}
\def\bSigma{{\boldsymbol{\Sigma}}}

\def\bchi{{\boldsymbol{\chi}}}
\def\bzero{{\boldsymbol{0}}}
\def\bdone{{\boldsymbol{1}}}
\def\bphi{{\boldsymbol{\phi}}}
\begin{document}

\title{Kernel-based multi-marker tests of association based on the accelerated failure time model
}

\titlerunning{Multi-Marker tests for the AFT model}        

\author{Chenxi Li         \and
        Di Wu \and Qing Lu 
}


\institute{C. Li \and D. Wu \at
              Department of Epidemiology and Biostatistics, Michigan State University, 909 Wilson Rd., East Lansing, MI 48824, USA \\
              \email{cli@msu.edu}\\ \email{wudi14@msu.edu}\\
            \and
            Q. Lu \at Department of Biostatistics, University of Florida, 2004 Mowry Road, Gainesville, FL 32611, USA \\
              \email{lucienq@ufl.edu}           
}


\maketitle

\begin{abstract}
Kernel-based multi-marker tests for survival outcomes use primarily the Cox model to adjust for covariates. The proportional hazards assumption made by the Cox model could be unrealistic, especially in the long-term follow-up. We develop a suite of novel multi-marker survival tests for genetic association based on the accelerated failure time model, which is a popular alternative to the Cox model due to its direct physical interpretation. The tests are based on the asymptotic distributions of their test statistics and are thus computationally efficient. The association tests can account for the heterogeneity of genetic effects across sub-populations/individuals to increase the power. All the new tests can deal with competing risks and left truncation. Moreover, we develop small-sample corrections to the tests to improve their accuracy under small samples. Extensive numerical experiments show that the new tests perform very well in various scenarios. An application to a genetic dataset of Alzheimer's disease illustrates the tests' practical utility.
\end{abstract}

\section{Introduction}\label{sec:intro}
Multi-marker tests have been popular for genome-wide association studies (GWAS) and transcriptomic profiling since the seminal paper on sequence kernel association test (SKAT) \citep{WuLe} was published. By testing the joint effect of genetic markers in a knowledge-based region (e.g., a gene region or a biological pathway), multi-marker tests aggregate the association signals and reduce the multiple testing burden as opposed to single-marker tests, thereby improving the power for association discovery. In addition, most multi-marker tests are kernel-based, which account for inter-marker correlations and thus have higher power compared to the regular tests for testing multiple markers, e.g., F-tests and likelihood-based tests (i.e., Wald, score and likelihood ratio tests).

Although there has been a rich literature of multi-marker tests for quantitative and binary traits, e.g., \citet{WuLe}, \citet{LeEm} and \citet{IoLe} to name a few,
 the field of multi-marker tests for censored survival outcomes is far less developed, primarily in four aspects. First, the types of covariate-adjustment models used by the existing multi-marker survival tests are limited. Most of the existing tests are based on the Cox model, including \citet{Goeman2005}, \citet{CaToLi}, \citet{Chen2014} and \citet{Lietal2020}. Only three works used non-Cox models, which are \citet{SiCa} for the accelerated failure time (AFT) model and \citet{Tzeng2014} and \citet{Wuetal2021} for linear transformation models. Misspecifying the covariate-adjustment model will lead to an incorrect null distribution for a multi-marker test, hampering the gene discovery process.
 Second, all the existing multi-marker survival tests apply only to time-to-event outcomes, while other types of survival phenotypes, e.g., competing risks and recurrent events, are not uncommon in genetic studies of human diseases. Third, all the existing tests except \citet{Goeman2005}, \citet{Chen2014} and \citet{Tzeng2014} are not valid in the general situation where the adjustment covariates are correlated with the genetic markers under testing; see \citet{Lietal2020} for relevant discussion and simulations. The tests of \citet{Lietal2020} can only adjust for linear confounding, namely the genetic markers are linearly correlated with the confounders. Fourth, many existing tests, including \citet{Goeman2005}, \citet{SiCa} and \citet{Chen2014},  are not accurate in terms of the null distribution of p-value under small or even modest sample sizes, as shown by \citet{Lietal2020}.

 In this article, we propose a set of multi-marker association tests for survival outcomes based on the accelerated failure time model, a popular alternative to the Cox model in survival analysis due to its direct physical interpretation \citep{Nancy}.  The proposed association tests can account for possible genetic heterogeneity (i.e., the genetic effect varies across sub-populations or individuals) to improve power. Compared to the existing AFT multi-marker test\citep{SiCa}, besides being able to account for genetic heterogeneity, our tests do not use sampling to compute the p-values, can deal with competing risks and left truncation, can adjust for confounding regardless of the relationship between the markers and the confounders, and most importantly are much more accurate under small and modest sample sizes.

 Our methods were motivated by the data from the Rush Memory and Aging Project and the Religious Orders Study (ROSMAP) \citep{ROSMAP}. These two studies are both ongoing cohort studies of aging and Alzheimer's disease. Both studies have over 20 years of follow-up and together generated genome-wide data  for over 1,600 subjects. So the ROSMAP data are a great resource for studying genetic risk factors for incident Alzheimer's disease. Nonetheless, the Cox model may not fit the time-to-AD data of ROSMAP given the long follow-up, and the genetic analysis of time-to-AD data needs to account for the competing risk of death without AD and the left truncation of survival outcome if the time scale is age. These considerations motivated us to develop the methods of this paper.

The rest of the paper is organized as follows. Section \ref{Methods} describes the proposed tests. The proof of their asymptotic null distributions is deferred to Appendix A.
Section \ref{sim} shows extensive simulations to evaluate the finite-sample performance of the methods. Section \ref{app} presents an application of the new tests to a gene-based association analysis of age at Alzheimer's disease onset with the ROSMAP data. The paper concludes with some discussion on future research directions in Section \ref{dis}.

\section{Methods}\label{Methods}
We develop the new association tests just for competing risks data under left truncation, since regular survival data with/without left truncation will be special cases where there is only one failure cause.

\subsection{Association tests}\label{sec:assoc_tests}
Consider a cohort study where participants have not experienced any competing risk events at baseline. The sample size is denoted by $n$. The observed survival data of a subject are $(A,\widetilde{T},\Delta,\epsilon\Delta)$, where $\widetilde{T}=\min(T,C)$, $T$ is the time to failure, $C$ is the time from the time origin for $T$ (the time point at which $T=0$) to censoring, $\Delta=I(T\le C)$, $\epsilon$ is the failure cause, and $A$ is the left truncation time, namely the time from the time origin for $T$ to the time point when the subject enters the follow-up, e.g., age at study entry in an analysis of age at onset of a disease. Suppose that there are $J$ failure causes, denoted by $1,\ldots,J$. We are interested in testing the effect of a set of genetic markers $\bG\equiv(G_1,\ldots,G_p)^T$, e.g., in a gene or biological pathway, on Cause 1, where $G_i$ denotes the value of the $i$-th marker $(i=1,\ldots,p)$, e.g., the number of minor alleles at the $i$-th SNP. We develop two association tests to accomplish this objective. One of the tests considers the possibility that the effect of $\bG$ varies across individuals or certain sub-populations, e.g., different genome profiles, sexes or races. The population structure is either explicit, which can be indicated by a vector of observable variables,  $\bX\equiv(X_1,\ldots,X_D)^T$ (e.g., race), or latent (e.g., sub-populations with different ancestry backgrounds) but inferable by a vector of  observable variables, also denoted by $\bX$ (e.g., a large number of SNPs from GWAS data). The other new association test does not consider genetic heterogeneity. In both tests, we adjust for covariates $\bZ\equiv(Z_1,\ldots, Z_q)^T$ to reduce confounding and/or increase power. $\bX$ is part of $\bZ$ when the sub-populations are explicit. We assume that the dyad of failure time and cause, the truncation time, and the residual censoring time $C-A$ are conditionally independent given $\bG$ and $\bZ$ (and $\bX$ if considering genetic heterogeneity).

The common null hypothesis of our association tests is that $\bG$ has no effect on the cause-specific hazard (CSH) of Cause 1 after adjusting for $\bZ$ (in any sub-population if considering genetic heterogeneity). We assume that, under the null, the CSH of Cause 1 given $\bZ$ follows an AFT model:
\bq\label{null_model}
\lambda(t;\bZ)=\lambda_0(te^{-\bbeta^T\bZ})e^{-\bbeta^T\bZ},
\eq
where $\lambda_0(\cdot)$ is an unspecified baseline CSH.

The association tests involve fitting the null model \eqref{null_model} to the data at first. Specifically, we estimate $\bbeta$ and $\Lambda_0(\cdot)\equiv\int_0^\cdot\lambda_0(s)ds$ by applying the rank-based estimation method \citep{ChXu} to a (working) AFT model:
\bq\label{working_null_model}\log(T^*)=\bbeta^T\bZ+\varepsilon. \eq
The (working) response $T^*$ is subject to left truncation with  truncation time being $A$, which equals $\widetilde{T}$ if $\epsilon\Delta=1$, and is right censored at $\widetilde{T}$ otherwise. The random error $\varepsilon$ is independent of $\bZ$, and the hazard function of $\varepsilon$ is $\lambda_{\varepsilon}(s)=e^s\lambda_0(e^s)$. Following \citet{ChXu}, we define $e^a_i(\bbeta)=\log A_i-\bbeta^T\bZ_i$, $e_i(\bbeta)=\log \widetilde{T}_i-\bbeta^T\bZ_i$, $N_i(\bbeta,t)=I(e_i(\bbeta)\le t, \epsilon\Delta=1)$, $Y_i(\bbeta,t)=I(e_i(\bbeta)\ge t)$, $\nu_i(\bbeta,t)=I(e^a_i(\bbeta)< t)$. The quantities $N_i(\bbeta,t)$ and $\nu_i(\bbeta,t)Y_i(\bbeta,t)$ are respectively the counting and the at-risk processes of subject $i$ on the transformed time scale of $\log(T^*)-\bbeta^T\bZ$.   Given $\bbeta$,  $\Lambda_\varepsilon(t)\equiv\int_{-\infty}^t\lambda_\varepsilon(s)ds$ is estimated by a Nelson-Aalen type estimator \citep{ChXu}, \bq\label{breslow_est}\widehat{\Lambda}_\varepsilon(\bbeta,t)=\int_{-\infty}^t\{\sum_{i=1}^n\nu_i(\bbeta,s)Y_i(\bbeta,s)\}^{-1}\sum_{i=1}^ndN_i(\bbeta,s).\eq
The regression coefficients, $\bbeta$, are estimated from the rank-based estimating equation,  Eq. (5) in \citet{ChXu}, with log-rank weights. The resulting estimator is denoted by $\widehat{\bbeta}$.

Define $M_i(\Lambda_\varepsilon,\bbeta)=N_i(\bbeta,\infty)-\int_{-\infty}^\infty\nu_i(\bbeta,t)Y_i(\bbeta,t)d\Lambda_\varepsilon(t)$, $M_i=M_i(\Lambda_\varepsilon,\bbeta)$, $\widehat{M}_i=M_i(\widehat{\Lambda}_\varepsilon(\bbeta,\cdot),\bbeta)$, and $\widetilde{M}_i=M_i(\widehat{\Lambda}_\varepsilon(\widehat{\bbeta},\cdot),\widehat{\bbeta})$. The quantity $\widetilde{M}_i$ can be viewed as a martingale residual on the transformed time scale of $\log(T^*)-\bbeta^T\bZ$ since the process $M_i(t)=N_i(\bbeta,t)-\int_{-\infty}^t\nu_i(\bbeta,u)Y_i(\bbeta,u)d\Lambda_\varepsilon(u)$ is a martingale when the null hypothesis of no association is true. The proposed association test that does not consider genetic heterogeneity is based on the test statistic,
\bq \label{assoc_test_stat} R\equiv\widetilde{\bM}^T\bK\widetilde{\bM},\eq
where $\widetilde{\bM}=(\widetilde{M}_1,\ldots,\widetilde{M}_n)^T$, $\bK=\{K(\bG_i,\bG_j)\}_{n\times n}$, and $K(\cdot,\cdot)$ is a Mercer kernel function \citep[][p. 35]{Herb}. Note that
\bq R=\sum_{i=1}^n\sum_{j=1}^n K(\bG_i,\bG_j)\widetilde{M}_i\widetilde{M}_j.\eq The quantity $\widetilde{M}_i\widetilde{M}_j$ can be viewed as a phenotype similarity between subjects $i$ and $j$ adjusted for $\bZ$. The quantity $K(\bG_i,\bG_j)$ equals the cross product of $\bphi(\bG_i)$ and $\bphi(\bG_j)$, the features obtained by mapping $\bG_i$ and $\bG_j$ respectively to the feature space induced by $K(\cdot,\cdot)$ \citep[][Chap. 2]{Herb}, and thus can be viewed as a genetic similarity between subjects $i$ and $j$. When $\bG$ has an effect on the CSH of Cause 1 while controlling for $\bZ$, the phenotype similarity $\widetilde{M}_i\widetilde{M}_j$ is concordant with the genetic similarity $K(\bG_i,\bG_j)$. In other words, larger (smaller) $\widetilde{M}_i\widetilde{M}_j$'s correspond to larger (smaller) $K(\bG_i,\bG_j)$'s, leading to a large value of $R$.

The choice of $K(\cdot,\cdot)$ depends on the expected effect form of $\bG$ in a way that the induced feature space by $K(\cdot,\cdot)$ should match that form. For example, if the effect of $\bG$ is expected to be linear, we use the linear kernel, a.k.a. cross-product kernel, $K(\bG_i,\bG_j)=\bG_i^T\bG_j$, for which the induced feature space is the input space $\{\bG\}$. If  $\bG$ is a vector of SNP covariates that are expected to have monotone but nonlinear  effects, we use the IBS kernel, $K(\bG_i,\bG_j)=\sum_{k=1}^p(2-|G_{i,k}-G_{j,k}|)/(2p)$. If $\bG$ is a vector of gene expression covariates that are expected to have nonlinear and/or interactive effects, we use a polynomial kernel, $K(\bG_i,\bG_j)=(\rho+\bG_i^T\bG_j)^d$, or the Gaussian kernel, $K(\bG_i,\bG_j)=\exp(-\rho\|\bG_i-\bG_j\|^2)$, where $d$ is a specified positive integer and $\rho$ is a specified positive constant. As suggested by \citet{WeLu}, a universal genetic similarity kernel could be the Laplacian kernel, $K(\bG_i,\bG_j)=\exp(-\sum_{k=1}^pw_k|G_{ik}-G_{jk}|/\Upsilon)$, where $G_{ik}$ can be discrete or continuous variables, $\Upsilon=\sum_{k=1}^pw_k $, and $w_k$ is the reciprocal of the sample standard deviation of $G_k$. This kernel is particularly useful for association mapping from sequencing reads, which involves many rare variants.

Suppose that the rank of $\bK$ is $m$. Then $\bK=\bE_1^T\bE_1$ for some $m\times n$ matrix $\bE_1$. In Appendix A, we derive an estimator of $Cov(\bE_1\widetilde{\bM})$ under the null hypothesis and a large sample size. Denote the eigenvalues of this covariance matrix estimator by $\lambda_1,\ldots,\lambda_m$. It is shown in Appendix A that the large-sample null distribution of $R$ is approximately $\sum_{j=1}^m\lambda_j\chi_{1j}^2,$
where $\chi_{1j}^2$'s are independent chi-square variables with degree 1. Based on this distribution, we compute the p-value, $P(R\ge R_{\mbox{obs}})$, using Davies' method \citep{Davies}.

The proposed association test that considers genetic heterogeneity is based on the test statistic, \bq R_{\mbox{het}}\equiv\widetilde{\bM}^T\bW\widetilde{\bM},\eq
where $\bW=(\bJ+\bH)\cdot\bK$, $\bJ=\{1\}_{n\times n}$, $\bH=\{H(\bX_i,\bX_j)\}_{n\times n}$, $H(\cdot,\cdot)$ is a kernel function measuring the sub-population similarity, and $\cdot$ represents the Hadamard product. The quantity $\bW$ can be viewed as a heterogeneity-weighted genetic similarity matrix. Note that $R_{\mbox{het}}=\widetilde{\bM}^T \bK\widetilde{\bM}+\widetilde{\bM}^T(\bH\cdot \bK)\widetilde{\bM}$. Thus it can be thought to simultaneously test the main effect of $\bG$ (through $\widetilde{\bM}^T \bK\widetilde{\bM}$) and its interaction with the sub-population (through $\widetilde{\bM}^T (\bH\cdot\bK)\widetilde{\bM}$).

The choice of $H(\cdot,\cdot)$ depends on the type of $\bX$. If $\bX$ is a set of dummy variables coding explicit sub-populations, e.g. different sexes, we can use the identity kernel $H(\bX_i,\bX_j)=I(\bX_i=\bX_j)$. If $\bX$ is a set of SNPs, we can choose the IBS kernel for $H(\bX_i,\bX_j)$. If $\bX$ is a set of continuous variables, the Gaussian kernel can be used.

The approximate null distribution of $R_{\mbox{het}}$ can be derived the same way as for $R$, except that $\bK$ is replaced by $\bW$ in the derivation. Based on this distribution, we compute the p-value, $P(R_{\mbox{het}}\ge R_{\mbox{het}}^{\mbox{obs}})$, using Davies' method \citep{Davies}.

The derivations of the large sample null distributions of $R$ and $R_{\mbox{het}}$ do not require any assumption about the relationship between $\bG$ and $\bZ$. Therefore, the two association tests can adjust for confounding regardless of the relationship between the markers and the confounders. This is a desirable property for (epi)genetic association tests and differential expression tests, because confounding is ubiquitous in those data analyses and usually has an unknown form.
Some early multi-marker survival tests cannot adjust for any confounding, e.g. \citet{CaToLi} and \citet{SiCa}, as shown in \citet{Lietal2020}. The multi-marker survival tests proposed by \citet{Lietal2020} can only adjust for linear confounding, namely, $\bG=\ba+\bB^T\bZ_c+\bltre$, where $\bZ_c$ is a vector of confounders, $\ba$ and $\bB$ are respectively a constant vector and a constant matrix, and $\bltre$ is a zero-mean random error vector that is independent of $\bZ_c$. In genetic association analyses, the markers and the confounders are usually minor allele counts of SNPs and the top few principal components of the genome-wide genotype data, respectively, and so the above linear model between $\bG$ and $\bZ_c$ cannot hold because the conditional variance of $\bG$ given $\bZ_c$ depends on the conditional mean. In differential gene expression analyses, confounding usually does not have the above linear form either. For example, $\bZ_c$ often includes age of the subject, and there is evidence that age affects both mean and variance of gene expression \citep{vinuela2018age}.

\subsection{Small-sample corrections to the proposed tests}\label{sec:smallsample}
When the sample size is relatively small in consideration of the number of genetic markers, the asymptotic null distributions of the test statistics are not accurate enough to approximate their finite-sample null distributions and the proposed tests tend to be conservative, as shown in the simulations. To address this issue, we  develop a small-sample correction strategy for the proposed tests. Specifically, we will change the test statistics $R$ and $R_{\mbox{het}}$ to $R^c\equiv\widetilde{\bM}^T\bK\widetilde{\bM}/{\widetilde{\bM}^T\widetilde{\bM}}$ and $R_{\mbox{het}}^c\equiv\widetilde{\bM}^T\bW\widetilde{\bM}/{\widetilde{\bM}^T\widetilde{\bM}}$ respectively. These changes were motivated by a similar correction strategy proposed for the kernel machine based association tests for quantitative traits \citep{Chenetal2016}. The p-value, $P(R^c\ge R^c_{\mbox{obs}})$, equals $P(\widetilde{\bM}^T(\bK-R^c_{\mbox{obs}}\bI_n)\widetilde{\bM}\ge 0)$, where $\bI_n$ is an $n\times n$ identity matrix. The large-sample null distribution of $\widetilde{\bM}^T(\bK-R^c_{\mbox{obs}}\bI_n)\widetilde{\bM}$ can be obtained following the derivation for $R$, which is also a linear combination of independent chi-square variables with degree 1. So we use Davies' method \citep{Davies} to compute the p-value. The p-value, $P(R_{\mbox{het}}^c\ge R^c_{\mbox{het,obs}})$, can be computed similarly.

\section{Numerical Experiments}\label{sim}

We performed Monte Carlo simulations to assess the finite-sample performance of the proposed tests with competing risk data under left truncation. In all the simulation scenarios, unless otherwise specified, two competing risks, two sample sizes, $n=400$ and $n=500$, and two marker-set sizes, $p=20$ and $p=25$, were considered. The competing risk data were generated from the AFT models specified later in specific scenarios, following the steps in Section 3.2 of \citet{BeyersJan}. The left truncation time was generated from $U(0,1)$, and the residual censoring time (censoring time since truncation) was generated from $\exp(0.1)$. The failure due to Cause 1 is of interest. When considering adjustment covariates, a binary covariate $Z_1 \sim Bernoulli(0.5)$ and a continuous covariate $Z_2 \sim U(0,2)$ were generated for each subject. The genetic markers under testing were SNPs, except in the scenario of confounding where the genetic markers were gene expression values. We generate SNP covariates by sampling from the genotype data of the 1000 genomes project (phase 3) \citep{1000G}. In all the simulation scenarios, 1000 Monte Carlo samples were generated, and the significance level of a test was set at 0.05, unless otherwise specified.

In Appendix B, we provided additional simulations to assess the performance of the association test considering genetic heterogeneity when the sub-populations are latent, simulations to assess the robustness of our association test against model misspecification,  simulations to compare the performances of our association test and the Cox model-based kernel association test (coxKM) \citep{CaToLi} with data generated from AFT models, and simulations to measure the runtimes of the proposed tests.

\subsection{Testing genetic association in the absence of genetic heterogeneity}
\label{sec3.1}

In this series of simulations, the performance of the test $R$ was assessed in detecting the association between a set of genetic markers and the failure due to Cause 1 in the absence of genetic heterogeneity. The hazard function of Cause 1 followed an AFT model,
\begin{equation}
    \lambda_1 (t|\bZ,\bG) = \lambda_0(t\cdot \exp \{-\sum_{j=1}^p \beta_j G_{j} - \sum_{k=1}^2 0.1 Z_{k}\}) \exp \{-\sum_{j=1}^p \beta_j G_{j} - \sum_{k=1}^2 0.1 Z_{k}\},
\label{aftmod11}
\end{equation}
and the hazard of Cause 2 followed another AFT model,
\begin{equation}
    \lambda_2 (t|\bZ,\bG) = \lambda_0(t \cdot \exp \{-\sum_{j=1}^p \alpha_j G_{j} - \sum_{k=1}^2 0.2 Z_{k}\}) \exp \{-\sum_{j=1}^p \alpha_j G_{j} - \sum_{k=1}^2 0.2 Z_{k}\},
\label{aftmod12}
\end{equation}
where the values of $\beta_j$'s and $\alpha_j$'s varied depending on the simulation scenario. The baseline hazard function was  $\lambda_0(x) = x^2 + x$.

\subsubsection{Empirical size and power of the test $R$ under no confounding}
\label{3.1.1}

In this simulation, we investigated the performance of the test $R$ in the absence of genetic heterogeneity and confounding effects under various $n$'s and $p$'s. For comparison, we also investigated the performance of $R_{\mbox{het}}$ in the same settings as for $R$. The genetic markers under testing were SNPs. We set the regression coefficients of $\bG$ in (\ref{aftmod11}) to be $\beta_j=0.08$ and $\beta_j=0$ ($j=1,\ldots,p$) in the power evaluation and the size assessment, respectively, and set the regression coefficients of $\bG$  in (\ref{aftmod12}) to be $\alpha_j=0.16$ ($j=1,\ldots,p$).  The IBS kernel was used to measure the genetic similarity in $R$ and $R_{\mbox{het}}$. The Gaussian kernel was used to measure the sub-population similarity in $R_{\mbox{het}}$, where the population structure was represented by the adjustment covariates. Table \ref{redo_main_text_table1} shows that the empirical sizes of both $R$ and $R_{\mbox{het}}$ are close to the nominal level under various $n$'s and $p$'s. The powers of $R$ and $R_{\mbox{het}}$ both increase with the sample size, and the former is a little higher than the latter due to the unnecessary accounting for heterogeneity by $R_{\mbox{het}}$ in this scenario.


\begin{table}[H]
\setlength\tabcolsep{2pt}
\caption{Empirical size and power comparison of $R$ and $R_{\mbox{het}}$ in testing genetic effects under covariate adjustment, left truncation and no genetic heterogeneity.}
\label{redo_main_text_table1}
\begin{tabular*}{\textwidth}{@{\extracolsep{\fill}} l *{3}{d{2.4}} }
\hline
 & \multicolumn{2}{c}{Empirical Size (Power)} \\ \cline{2-3}
 & \multicolumn{1}{c}{p=20, n=400} & \multicolumn{1}{c}{p=20, n=500} \\ \cline{2-3}
\multicolumn{1}{l}{$R$} & \multicolumn{1}{c}{0.044 (0.447)} & \multicolumn{1}{c}{0.052 (0.574)} \\
\multicolumn{1}{l}{$R_{het}$} & \multicolumn{1}{c}{0.045 (0.426)} & \multicolumn{1}{c}{0.049 (0.537)} \\ \cline{2-3}
 & \multicolumn{1}{c}{p=25, n=400} & \multicolumn{1}{c}{p=25, n=500} \\ \cline{2-3}
\multicolumn{1}{l}{$R$} & \multicolumn{1}{c}{0.046 (0.540)} & \multicolumn{1}{c}{0.042 (0.646)} \\
\multicolumn{1}{l}{$R_{het}$} & \multicolumn{1}{c}{0.042 (0.476)} & \multicolumn{1}{c}{0.046 (0.603)} \\
\hline
\end{tabular*}
\end{table}

\subsubsection{Empirical size and power of the test $R$ under quadratic confounding}
\label{3.1.2}

\begin{figure}[H]
\captionsetup[subfigure]{justification=centering}
\centering
\subcaptionbox{$n=400, p=20$}{\includegraphics[width=0.5\textwidth]{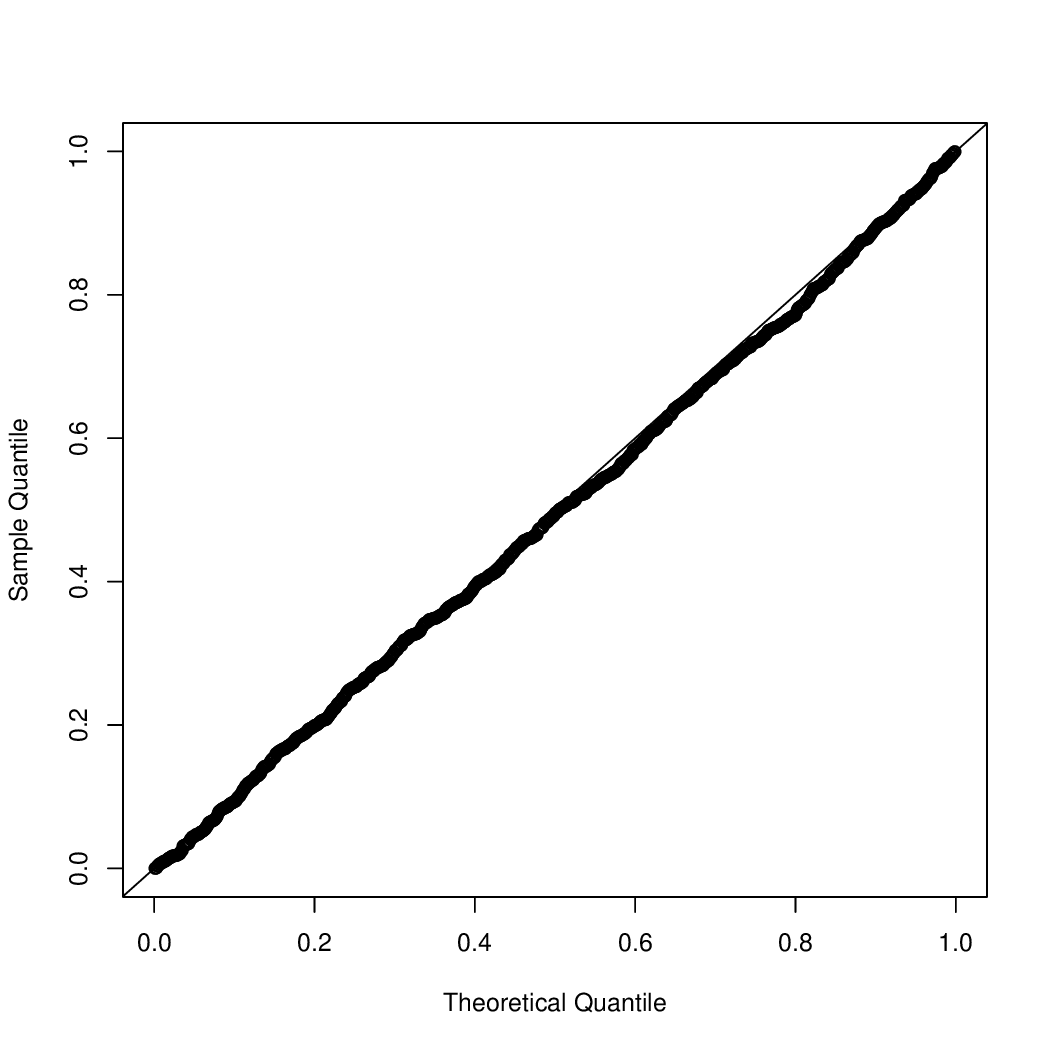}}%
\hfill 
\subcaptionbox{$n=400, p=25$}{\includegraphics[width=0.5\textwidth]{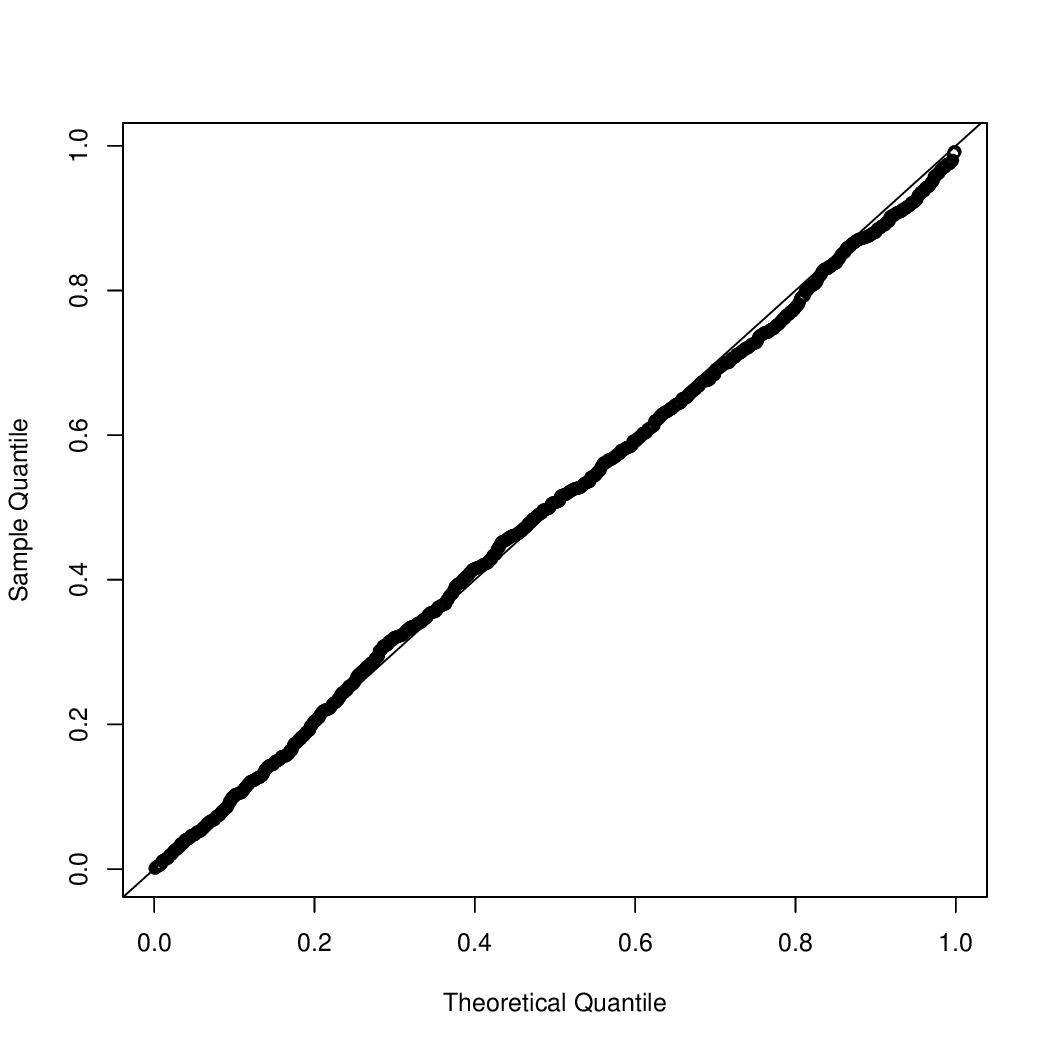}}%
\caption{Uniform Q-Q plots of the null p-value of $R$ under quadratic confounding with $n=400$ and $p=20$ or $25$.}
\label{R_confounding}
\end{figure}

In this simulation, the adjustment covariates $Z_1$ and $Z_2$ were confounders, and the genetic markers under testing were gene expressions. To simulate confounding effects, We assume $\bG_i = \mathbf{0.5}Z_{i1} + \mathbf{0.5}Z_{i2} + \mathbf{0.25}Z_{i1}^2 + \mathbf{0.25}Z_{i2}^2 + \mathbf{0.5}Z_{i1}Z_{i2} + \bde_{i}$, where $\bG_i=(G_{i1},\dots,G_{ip})^T$ were the expression levels of $p$ genes in subject $i$, $\mathbf{0.25}$ and $\mathbf{0.5}$ are $p$-dimensional vectors of 0.25's and 0.5's respectively, and $\bde_{i}=(e_{i1},\dots,e_{ip})^T$ follows a multivariate normal distribution with a zero mean and the covariance matrix being $\bSigma_{p\times p}=\{0.1^{|k-l|}\}$. The corresponding confounding effect of $Z_1$ and $Z_2$ is called quadratic confounding. As discussed in Section \ref{sec:assoc_tests}. many of the existing multi-marker survival tests, including \citet{CaToLi} and \citet{SiCa}, cannot adjust for confounding at all, and the tests of \citet{Lietal2020} can only adjust for linear confounding. So we use simulations under quadratic confounding to illustrate that our association test $R$ can adjust for confounding regardless of the relationship between the genetic markers and the confounders.
The regression coefficients of $\bG$ in \eqref{aftmod11} were set to be $\beta_j=0$ for the size assessment and $\beta_j=0.03$ for the power evaluation ($j=1,\ldots,p$). In the hazard function \eqref{aftmod12}, we set $\alpha_j=0.1$ for both the empirical size and power evaluations. We used the Gaussian kernel to measure the gene expression similarity in $R$. Figure \ref{R_confounding} shows that the p-value of $R$ under the null follows a $U[0,1]$ distribution when adjusting for confounders. Table \ref{hwv_quad_confound} shows that under  quadratic confounding, the empirical size of $R$ is still close to the nominal level, and the power of the test increases with the sample size.

\begin{table}[H]
\centering
\setlength\tabcolsep{36pt}
\caption{Empirical size and power of $R$ in testing genetic effects under quadratic confounding and left truncation.}
\label{hwv_quad_confound}
\begin{tabular}{@{\extracolsep{\fill}} c c }
\hline
  \multicolumn{2}{c}{Empirical Size (Power)} \\ \hline
   \multicolumn{1}{c}{p=20, n=400} & \multicolumn{1}{c}{p=20, n=500} \\ \hline
 \multicolumn{1}{c}{0.056 (0.771)} & \multicolumn{1}{c}{0.052 (0.863)} \\ \hline
  \multicolumn{1}{c}{p=25, n=400} & \multicolumn{1}{c}{p=25, n=500} \\ \hline
\multicolumn{1}{c}{0.046 (0.816)} & \multicolumn{1}{c}{0.052 (0.890)} \\
\hline
\end{tabular}
\end{table}

\subsection{Testing genetic association in the presence of genetic heterogeneity}
\label{sec3.2}

In this simulation, we investigated the empirical size and power of $R_{\mbox{het}}$ for testing the joint effect of a SNP set in the presence of genetic heterogeneity across two observable sub-populations with equal proportions. For comparison, we also investigated the performance of $R$ in the same settings. Since we generated the SNPs by sampling from the 1000 Genomes data set, we let the two observable sub-populations be males and females and associated a sampled SNP set with the sub-population indicated by the sex of the subject from whom the SNP set was obtained.

The survival time of Subject $i$ $(i=1,\ldots,n)$ was generated from the following AFT models for the cause-specific hazard functions,
\begin{equation}
    \lambda_1 (t|Z_i,\bG_i) = \lambda_0(t\cdot \exp \{-\sum_{j=1}^p (\beta_0 + \beta_1 Z_i)G_{ij} - 0.5 Z_i\}) \exp \{-\sum_{j=1}^p (\beta_0 + \beta_1 Z_i)G_{ij} - 0.5 Z_i\},
\label{aftmod21}
\end{equation}

\begin{equation}
    \lambda_2 (t|Z_i,\bG_i) = \lambda_0(t\cdot \exp \{-\sum_{j=1}^p 0.2 G_{ij} -  Z_i\}) \exp \{-\sum_{j=1}^p 0.2 G_{ij} -  Z_i\},
\label{aftmod22}
\end{equation}
where $Z_i $ is the sex variable included in the 1000 Genomes data set. To assess the size and the power of $R_{\mbox{het}}$, we set $\beta_0 = 0$ and $\beta_0 = 0.002$, respectively. We also set $\beta_1=0$ in the size assessment and  $\beta_1 = 0.1$ and $ 0.2$ in the power assessment, with the larger $\beta_1$ representing the stronger genetic heterogeneity. The IBS kernel was used to measure the genetic similarity in $R_{\mbox{het}}$ and $R$, and the identity kernel $I(Z_i=Z_j)$ was used to measure the sub-population similarity in $R_{\mbox{het}}$. Table \ref{hwv_obs2pop} shows that the empirical sizes of  $R_{\mbox{het}}$ and  $R$ are both around the nominal level. Table \ref{hwv_obs2pop} also shows that $R_{\mbox{het}}$ has higher power than  $R$ by accounting for the genetic heterogeneity across the two observable sub-populations. As the heterogeneity size (measured by $\beta_1$) increases, the power advantage of $R_{\mbox{het}}$ against $R$ gets more obvious.

\begin{table}[H]
\setlength\tabcolsep{2pt}
\caption{Empirical sizes and powers of $R_{\mbox{het}}$ and $R$ in testing genetic association under genetic heterogeneity across two observable sub-populations and left truncation.}
\label{hwv_obs2pop}
\begin{tabular*}{\textwidth}{@{\extracolsep{\fill}} l *{7}{d{2.4}} }
\hline
 & \multicolumn{6}{c}{Size/Power} \\ \cline{2-7}
 & \multicolumn{3}{c}{p=20, n=400} & \multicolumn{3}{c}{p=20, n=500} \\ \cline{2-4}  \cline{5-7}
\multicolumn{1}{c}{($\beta_0$, $\beta_1$)} & \multicolumn{1}{c}{(0, 0)} & \multicolumn{1}{c}{(0.002, 0.1)} & \multicolumn{1}{c}{(0.002, 0.2)} & \multicolumn{1}{c}{(0, 0)} & \multicolumn{1}{c}{(0.002, 0.1)} & \multicolumn{1}{c}{(0.002, 0.2)} \\ \cline{2-4}  \cline{5-7}
\multicolumn{1}{l}{$R$} & \multicolumn{1}{c}{0.042} & \multicolumn{1}{c}{0.161} & \multicolumn{1}{c}{0.336} &
\multicolumn{1}{c}{0.052} & \multicolumn{1}{c}{0.196} & \multicolumn{1}{c}{0.426} \\
\multicolumn{1}{l}{$R_{\mbox{het}}$} & \multicolumn{1}{c}{0.042} & \multicolumn{1}{c}{0.207} & \multicolumn{1}{c}{0.435} &
\multicolumn{1}{c}{0.047} & \multicolumn{1}{c}{0.268} & \multicolumn{1}{c}{0.561} \\ \cline{2-4}  \cline{5-7}
 & \multicolumn{3}{c}{p=25, n=400} & \multicolumn{3}{c}{p=25, n=500} \\ \cline{2-4} \cline{5-7}
\multicolumn{1}{c}{($\beta_0$, $\beta_1$)} & \multicolumn{1}{c}{(0, 0)} & \multicolumn{1}{c}{(0.002, 0.1)} & \multicolumn{1}{c}{(0.002, 0.2)} & \multicolumn{1}{c}{(0, 0)} & \multicolumn{1}{c}{(0.002, 0.1)} & \multicolumn{1}{c}{(0.002, 0.2)} \\ \cline{2-4}  \cline{5-7}
\multicolumn{1}{l}{$R$} & \multicolumn{1}{c}{0.047} & \multicolumn{1}{c}{0.175} & \multicolumn{1}{c}{0.288} &
\multicolumn{1}{c}{0.050} & \multicolumn{1}{c}{0.263} & \multicolumn{1}{c}{0.347} \\
\multicolumn{1}{l}{$R_{\mbox{het}}$} & \multicolumn{1}{c}{0.041} & \multicolumn{1}{c}{0.231} & \multicolumn{1}{c}{0.355} &
\multicolumn{1}{c}{0.045} & \multicolumn{1}{c}{0.327} & \multicolumn{1}{c}{0.466} \\
\hline
\end{tabular*}
\end{table}

In Appendix B, we showed via simulations that $R_{\mbox{het}}$ also performed well
in the presence of genetic heterogeneity across latent sub-populations and across individual genome profiles.

\subsection{Small-sample adjustment}
\label{sec:smallsample_sim}

In this series of simulations, we assess the performances of the small-sample corrected tests proposed in Section \ref{sec:smallsample} by comparing the empirical sizes and powers of: 1) $R^C$ and $R$ in testing genetic association in the absence of genetic heterogeneity and 2) $R^c_{\mbox{het}}$ and $R_{\mbox{het}}$ in testing genetic association in the presence of genetic heterogeneity across two observable sub-populations. We use small $n$ relative to $p$.

\subsubsection{Empirical sizes and powers of $R$ and $R^c$}
\label{sec:smallsample_sim.1}
\begin{figure}[H]
\captionsetup[subfigure]{justification=centering}
\centering
\subcaptionbox{$R$}{\includegraphics[width=0.5\textwidth]{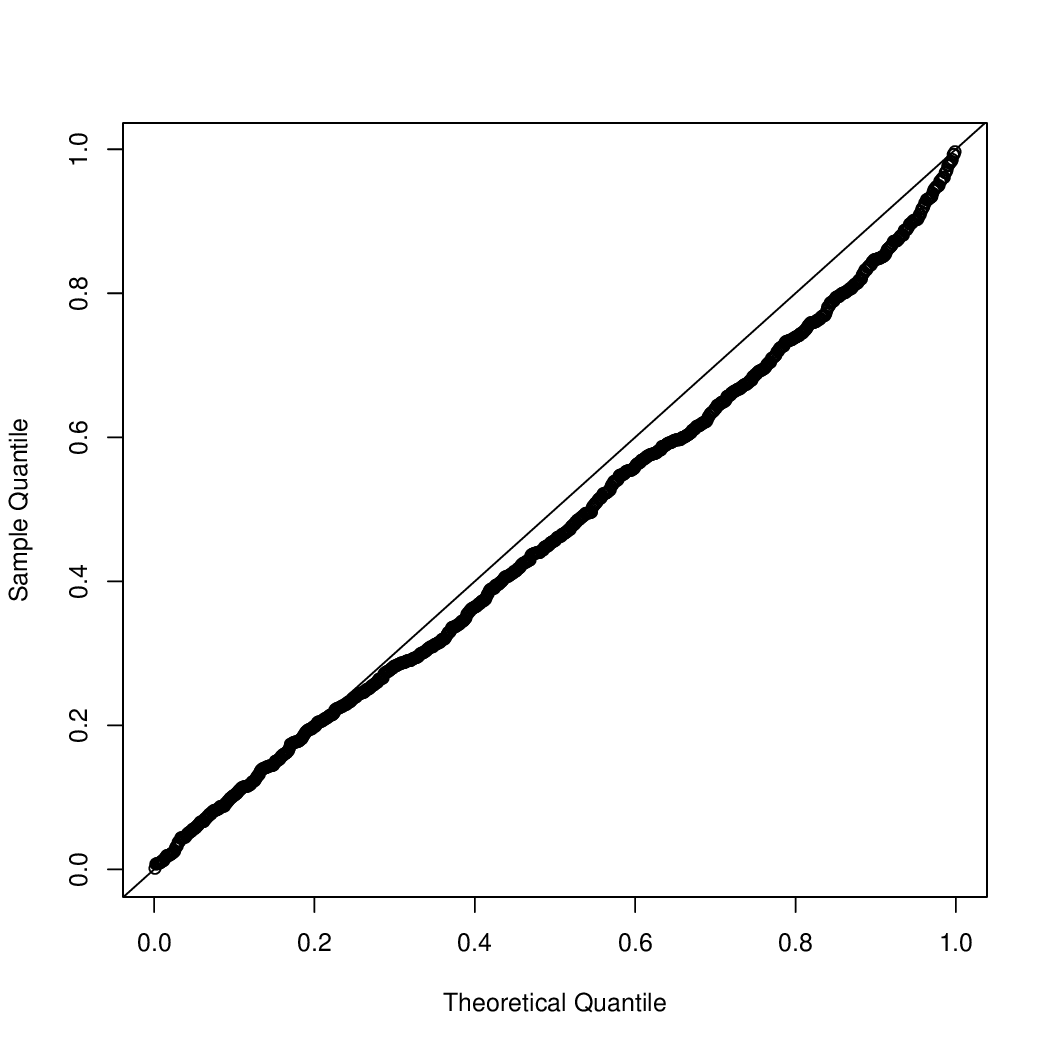}}%
\hfill 
\subcaptionbox{$R^c$}{\includegraphics[width=0.5\textwidth]{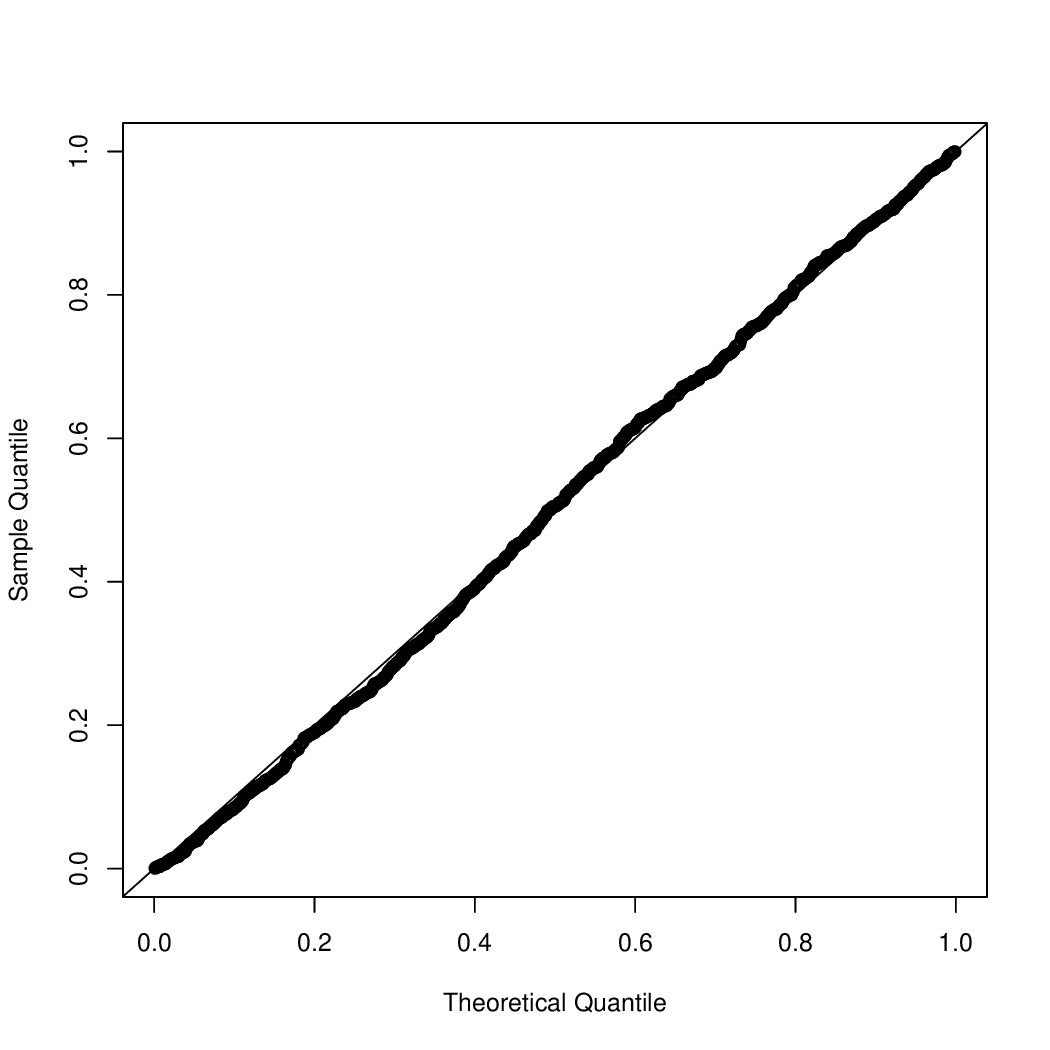}}%
\caption{Uniform Q-Q plots of the null p-value of $R$ and $R^c$ in testing genetic association under no genetic heterogeneity with $n=100$ and $p=15$.}
\label{smalln_nohet}
\end{figure}

In this simulation, we investigated the performances of $R$ and $R^c$ in testing genetic association in the absence of genetic heterogeneity. The simulation setting is similar to the previous setting for testing association under no genetic heterogeneity. The setting changes are the following. We decreased the sample size to $n=100$ and set $p=15$. The cause-specific hazard functions were
\begin{equation}
    \lambda_1 (t|\bZ,\bG) = \lambda_0(t\cdot \exp \{-\sum_{j=1}^p \beta_j G_{j} - \sum_{k=1}^2 0.1 Z_{k}\}) \exp \{-\sum_{j=1}^p \beta_j G_{j} - \sum_{k=1}^2 0.1 Z_{k}\}
\label{aftmod11_for_smalln}
\end{equation}
and
\begin{equation}
    \lambda_2 (t|\bZ,\bG) = \lambda_0(t \cdot \exp \{-\sum_{j=1}^p 0.2 G_{j} - \sum_{k=1}^2 0.2 Z_{k}\}) \exp \{-\sum_{j=1}^p 0.2 G_{j} - \sum_{k=1}^2 0.2 Z_{k}\},
\label{aftmod12_for_smalln}
\end{equation}
where $\beta_j = 0.1$ and $\beta_j = 0$ $(j=1,\ldots,p)$ in the power evaluation and size assessment respectively.   The Uniform Q-Q plots in Figure \ref{smalln_nohet} show that the null distribution of $R^c$'s p-value is closer to $U(0,1)$ than that of $R$'s p-value under small samples. The comparison of the size and power of $R$ and $R^c$ is shown in Table \ref{wv_smalln}. The results indicate that even when the sample size is small, both of the tests control Type I error well, but the test $R^c$ is more powerful than $R$.

\begin{table}[H]
\setlength\tabcolsep{2pt}
\caption{Empirical sizes and powers of $R$ and $R^c$ under left truncation with $n=100$ and $p=15$.}
\label{wv_smalln}
\begin{tabular*}{\textwidth}{@{\extracolsep{\fill}} l *{2}{d{2.4}} }
\hline
 & \multicolumn{1}{c}{Empirical Size (power)} \\ \cline{2-2}
\multicolumn{1}{c}{$R$} & \multicolumn{1}{c}{0.042 (0.528)}  \\
\multicolumn{1}{c}{$R^c$} & \multicolumn{1}{c}{0.054 (0.663)} \\
\hline
\end{tabular*}
\end{table}

\subsubsection{Empirical sizes and powers of $R_{\mbox{het}}$ and $R^c_{\mbox{het}}$}
\label{sec:smallsample_sim.2}
\begin{figure}[H]
\captionsetup[subfigure]{justification=centering}
\centering
\subcaptionbox{$R_{\mbox{het}}$}{\includegraphics[width=0.5\textwidth]{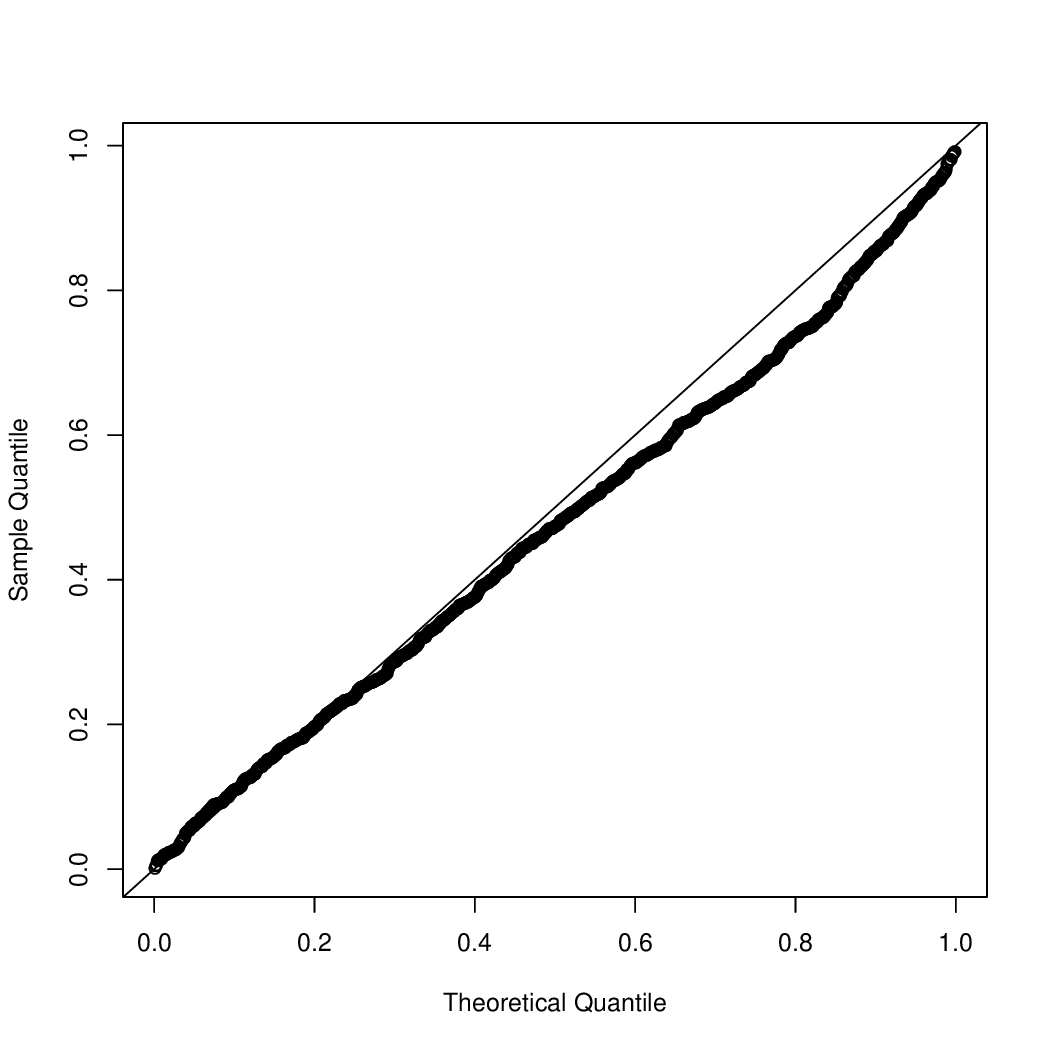}}%
\hfill 
\subcaptionbox{$R^c_{\mbox{het}}$}{\includegraphics[width=0.5\textwidth]{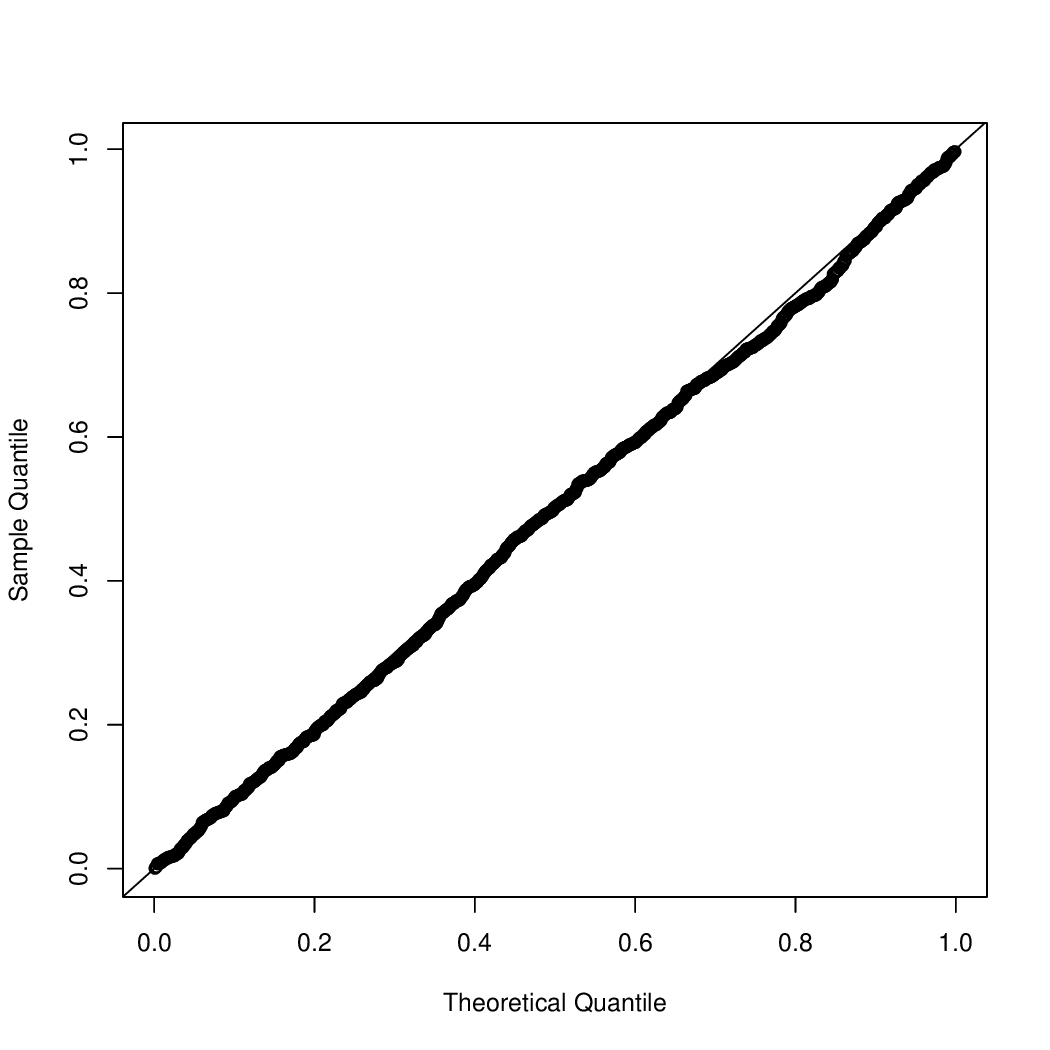 }}%
\caption{Uniform Q-Q plots of the null p-values of $R_{\mbox{het}}$ and $R^c_{\mbox{het}}$ considering genetic heterogeneity across two observable sub-populations, with $n=200$ and $p=10$.}
\label{smalln_het}
\end{figure}

In this simulation, we investigated the performances of  $R_{\mbox{het}}$ and $R^c_{\mbox{het}}$ in testing genetic association under genetic heterogeneity across two observable sub-populations. We set $n=200$ and $p=10$. The cause-specific hazard functions were:
\begin{equation}
    \lambda_1 (t|Z_i,\bG_i) = \lambda_0(t\cdot \exp \{-\sum_{j=1}^p (\beta_0 + \beta_1 Z_i)G_{ij} - 0.5 Z_i\}) \exp \{-\sum_{j=1}^p (\beta_0 + \beta_1 Z_i)G_{ij} - 0.5 Z_i\}
\label{aftmod21_for_small_n}
\end{equation}
and
\begin{equation}
    \lambda_2 (t|Z_i,\bG_i) = \lambda_0(t\cdot \exp \{-\sum_{j=1}^p 0.35 G_{ij} -  Z_i\}) \exp \{-\sum_{j=1}^p 0.35 G_{ij} - Z_i\},
\label{aftmod22_for_small_n}
\end{equation}
where $Z_i$ is the sex variable included in the 1000 Genomes (phase 3) data set. To assess the size and the power of $R_{\mbox{het}}$ and $R^c_{\mbox{het}}$, we set $\beta_0 = 0$ and $\beta_0 = 0.002$, respectively. We also set $\beta_1=0$ in the size assessment and  $\beta_1 = 0.15$ and $ 0.25$ in the power assessment, with the larger $\beta_1$ representing the stronger genetic heterogeneity. The Uniform Q-Q plots in Figure \ref{smalln_het} show that the null distribution of $R^c_{\mbox{het}}$'s p-value is closer to $U(0,1)$ than that of $R_{\mbox{het}}$'s p-value under small samples. Table \ref{hwv_smalln} shows that  both of the tests control Type I error well, but the test $R^c_{\mbox{het}}$ is more powerful than $R_{\mbox{het}}$.

\begin{table}[H]
\setlength\tabcolsep{2pt}
\caption{Empirical sizes and powers of $R_{\mbox{het}}$ and $R^c_{\mbox{het}}$ under genetic heterogeneity across two observable sub-populations and left truncation, with $n=200$ and $p=10$. }
\label{hwv_smalln}
\begin{tabular*}{\textwidth}{@{\extracolsep{\fill}} l *{4}{d{2.4}} }
\hline
 & \multicolumn{1}{c}{Empirical Size} & \multicolumn{2}{c}{Power} \\ \cline{2-4}
\multicolumn{1}{c}{$\beta_1$} & \multicolumn{1}{c}{0} & \multicolumn{1}{c}{0.15} & \multicolumn{1}{c}{0.25} \\ \cline{2-4}
\multicolumn{1}{c}{$R_{\mbox{het}}$} & \multicolumn{1}{c}{0.040} & \multicolumn{1}{c}{0.167} & \multicolumn{1}{c}{0.322}  \\
\multicolumn{1}{c}{$R^c_{\mbox{het}}$} & \multicolumn{1}{c}{0.051} & \multicolumn{1}{c}{0.211} & \multicolumn{1}{c}{0.381} \\
\hline
\end{tabular*}
\end{table}

\subsection{Empirical sizes of $R$ and $R_{\mbox{het}}$ under stringent p-value thresholds}
\label{sec_stringent_p_sim}

Genome-wide association studies with genotyping or sequencing data usually test the associations between hundreds of thousands of genetic variants and a phenotype, causing a severe multiple testing problem. Common approaches to address the multiple testing issue, such as the Bonferroni correction and the Benjamini-Hochberg procedure \citep{BeHo}, lead to stringent p-value thresholds when applied to those association analyses. In this simulation, we checked the sizes of the tests $R$, $R_{\mbox{het}}$, $R^c$, and $R^c_{\mbox{het}}$ under stringent p-value thresholds (i.e., those that are much smaller than 0.05). The simulation setting for $R$ and $R^c$ was the same as that for assessing $R$'s size and power under the 0.05 level. The simulation setting for $R_{\mbox{het}}$ and $R^c_{\mbox{het}}$ was the same as that for assessing $R_{\mbox{het}}$'s size and power in the presence of genetic heterogeneity across two observable sub-populations under the 0.05 level. In both scenarios, 500K Monte Carlo samples with $n=500$ and $p=20$ were generated to calculate the empirical sizes. Table \ref{hwv_stringentp} shows that the empirical sizes of $R^c$ and $R^c_{\mbox{het}}$ are very close to the stringent $p$-value thresholds. Those of $R$ and $R_{\mbox{het}}$ are close to the nominal level under the threshold of 0.05, but the two tests became conservative under the smaller thresholds, which was probably due to the relatively small sample size to the number of markers. When we increased $n$ to 1000, the empirical sizes of $R$ and $R_{\mbox{het}}$ got much closer to the nominal level under the stringent $p$-value thresholds (Table \ref{hwv_stringentp_largern}).
These results indicate that the proposed association tests, especially $R^c$ and  $R^c_{\mbox{het}}$, are suitable for large-scale genetic association analyses.

\begin{table}[H]
\setlength\tabcolsep{2pt}
\caption{Empirical sizes of $R$, $R_{\mbox{het}}$, $R^c$ and $R^c_{\mbox{het}}$ under stringent $p$-value thresholds and left truncation, with $n=500$ and $p=20$.}
\label{hwv_stringentp}
\begin{tabular*}{\textwidth}{@{\extracolsep{\fill}} l *{5}{d{2.4}} }
\hline
 & \multicolumn{4}{c}{Empirical Size} \\ \cline{2-5}
\multicolumn{1}{l}{Threshold} & \multicolumn{1}{c}{$R$} & \multicolumn{1}{c}{$R_{\mbox{het}}$} &
\multicolumn{1}{c}{$R^c$} & \multicolumn{1}{c}{$R^c_{\mbox{het}}$}
\\ \hline
\multicolumn{1}{l}{0.05} & \multicolumn{1}{l}{0.045} & \multicolumn{1}{l}{0.041} &
\multicolumn{1}{l}{0.051} &
\multicolumn{1}{l}{0.051} \\
\multicolumn{1}{l}{0.005} & \multicolumn{1}{l}{0.0037} & \multicolumn{1}{l}{0.0033} &
\multicolumn{1}{l}{0.0051} &
\multicolumn{1}{l}{0.0053} \\
\multicolumn{1}{l}{0.0005} & \multicolumn{1}{l}{0.00028} & \multicolumn{1}{l}{0.00023} &
\multicolumn{1}{l}{0.00052} &
\multicolumn{1}{l}{0.00058}\\
\multicolumn{1}{l}{0.00005} & \multicolumn{1}{l}{0.000026} & \multicolumn{1}{l}{0.000012} &
\multicolumn{1}{l}{0.000050} &
\multicolumn{1}{l}{0.000050}\\
\hline
\end{tabular*}
\end{table}

\begin{table}[H]
\setlength\tabcolsep{2pt}
\caption{Empirical sizes of $R$ and $R_{\mbox{het}}$ under stringent $p$-value thresholds and left truncation, with $n=1000$ and $p=20$.}
\label{hwv_stringentp_largern}
\begin{tabular*}{\textwidth}{@{\extracolsep{\fill}} l *{3}{d{2.4}} }
\hline
 & \multicolumn{2}{c}{Empirical Size} \\ \cline{2-3}
\multicolumn{1}{l}{Threshold} & \multicolumn{1}{c}{$R$} & \multicolumn{1}{c}{$R_{\mbox{het}}$} \\ \hline
\multicolumn{1}{l}{0.05} & \multicolumn{1}{l}{0.047} & \multicolumn{1}{l}{0.045} \\
\multicolumn{1}{l}{0.005} & \multicolumn{1}{l}{0.0041} & \multicolumn{1}{l}{0.0040} \\
\multicolumn{1}{l}{0.0005} & \multicolumn{1}{l}{0.00038} & \multicolumn{1}{l}{0.00038} \\
\multicolumn{1}{l}{0.00005} & \multicolumn{1}{l}{0.00038} & \multicolumn{1}{l}{0.00046} \\
\hline
\end{tabular*}
\end{table}

\section{A Real Application}\label{app}

We applied our tests $R^c$ and $R^c_{\mbox{het}}$ to the GWAS and Alzheimer's disease (AD) diagnosis data from two large longitudinal studies of aging and dementia, the Religious Orders Study (ROS) and the Rush Memory and Aging Project (MAP) \citep{ROSMAP}, collectively called ROSMAP. The goal of this real data analysis is to discover genes that are associated with age at AD onset. Death before AD onset is a competing risk. We excluded the subjects who had AD at the baseline visit from the analysis, leading to left truncation with the baseline age as the truncation time. Both the competing risk and the left truncation were accounted for in our analysis.

The GWAS dataset includes 1,679 subjects and 750,173 SNPs. After performing SNP-level quality control---removing SNPs with minor allele frequency (MAF)$<0.01$, Hardy–Weinberg equilibrium test's p-value$<10^{-6}$, or missing rate$>0.02$, 619,061 SNPs remained for the analysis. We then performed subject-level quality control to remove subjects with missing SNP genotype rate$>0.02$. After that, 1,618 subjects remained for the analysis.  The missing genotypes in the remaining genetic data were  imputed by IMPUTE v2.3.2 (\url{https://mathgen.stats.ox.ac.uk/impute/impute_v2.html#download}) with pre-phasing \citep{howie2012fast}. Then the 619,061 SNPs were grouped into gene-based SNP sets based on the human genome reference hg18 (i.e., located in or within 5K base pairs upstream/downstream of a gene). The grouping formed 21,285 genes along with the \textit{APOE} gene, which was coded as the count of \textit{APOE}-$\epsilon4$ alleles. Table \ref{snp_set_distribution} summarizes the distribution of  the SNP set size.
We performed a principal component analysis of the imputed GWAS data to obtain the first six principal components for adjusting for population stratification. The AD diagnosis dataset we used from the ROS and MAP studies was frozen in 2021 with a sample size of 3,675. It contains annual clinical diagnosis of AD since baseline. Out of the 3,675 subjects, 218 subjects had AD at baseline and thus were removed from the analysis. We treated the age at the first diagnosis of AD as the age at AD onset. We then merged the age at AD onset data with the processed GWAS data to generate the final analysis-ready data, which contains 1,440 subjects who have both the genetic data and the survival data. Among them, 540 subjects developed AD during the follow-up. In the analysis, to improve power and/or reduce confounding, we adjusted for the first six principal components from the GWAS data as well as sex, cohort (ROS or MAP) and education attainment (0: $\mbox{years of education}\le12$; 1: $12<\mbox{years of education}\le 16$; 2: $\mbox{years of education}>16$). We did not adjust for self-reported race because all but one of the 1440 subjects reported to be white.

\begin{table}[H]
\setlength\tabcolsep{1pt}
\caption{Percentiles of the SNP set size in ROSMAP genotype data.}
\label{snp_set_distribution}
\begin{tabular*}{\textwidth}{@{\extracolsep{\fill}} l *{12}{d{2.4}} }
\hline
 & 0\% & 10\% & 20\% & 30\% & 40\% & 50\% & 60\% & 70\% & 80\% & 90\% & 100\% \\
Percentile & 1 & 1 & 2 & 3 & 5 & 6 & 9 & 12 & 18 & 35 & 1434 \\
\hline
\end{tabular*}
\end{table}

Without considering genetic heterogeneity, we used the test $R^c$ to perform a genome-wide gene-based association scan by testing the effect of each of the 21,285 genes on age at AD onset. Four genetic similarity kernels were used: the IBS, linear, Laplacian and quadratic kernels. Controlling the false discover rate under 10\% by the Benjamini-Hochberg procedure \citep{BeHo}, \textit{APOE} and \textit{APOC1} appeared to be two significant genes no matter which genetic similarity kernel was used (Table \ref{realdataanalysis}). \textit{APOE}'s $p$-value ranged from 1.11E-16 to 1.34E-13, and \textit{APOC1}'s $p$-value ranged from 2.03E-10 to 5.01E-09. \textit{APOE} is an established susceptibility gene for AD, and \textit{APOC1} has also been reported to be an AD risk gene \citep[See, e.g.,][]{KULMINSKI2022122,Zhouetal2019}.

\begin{table}[H]
\footnotesize
\setlength{\tabcolsep}{9pt}
\renewcommand{\arraystretch}{1.25}
\caption{Top five genes discovered by $R^c$ and $R^c_{\mbox{het}}$ with the ROSMAP data. IBS, Lin, Lap and Quad stand for the IBS, linear, Laplacian and quadratic kernels, respectively. Various types of heterogeneity were considered, including no genetic heterogeneity (S1), heterogeneity between sexes (S2), heterogeneity across education attainment categories (S3), and heterogeneity across genetic backgrounds (S4). }
\label{realdataanalysis}
{\begin{tabular}{lllllll}
\hline
\multicolumn{1}{c}{Genetic} & \multicolumn{1}{c}{Scenario} & \multicolumn{5}{c}{Genes and p-values} \\
\multicolumn{1}{c}{similarity kernel} & \multicolumn{1}{c}{} & \multicolumn{5}{c}{} \\

\hline

 & \multicolumn{1}{c}{\multirow{2}{*}{S1}} & \multirow{2}*{\textbf{APOE}} & \multirow{2}*{\textbf{APOC1}} & \multirow{2}*{IGSF23} & PLEKHG5,& \multirow{2}*{TBCC} \\
 & & & & & TNFRSF25 &  \\
 & & \textbf{2.41E-14} & \textbf{4.61E-09} & 4.99E-05 & 7.91E-05 & 8.11E-05 \\ \cline{3-7}
\multicolumn{1}{c}{\multirow{4}{*}{IBS}} & \multicolumn{1}{c}{\multirow{2}*{S2}} & \multirow{2}*{\textbf{APOE}} & \multirow{2}*{\textbf{APOC1}} & PLEKHG5,& \multirow{2}*{IGSF23} & \multirow{2}*{TBCC} \\
& & & & TNFRSF25 & &  \\
&  & \textbf{8.04E-14} & \textbf{3.25E-09} & 5.64E-05 & 1.35E-04 & 1.42E-04 \\  \cline{3-7}
 & \multicolumn{1}{c}{\multirow{2}*{S3}} & \multirow{2}*{\textbf{APOE}} & \multirow{2}*{\textbf{APOC1}} & \multirow{2}*{IGSF23} & \multirow{2}*{TBCC} & PLEKHG5, \\
  & & & & & & TNFRSF25 \\
 & & \textbf{3.01E-14} & \textbf{2.53E-09} & 4.50E-05 & 4.84E-05 & 7.17E-05 \\  \cline{3-7}
 & \multicolumn{1}{c}{\multirow{2}*{S4}} & \multirow{2}*{\textbf{APOE}} & \multirow{2}*{\textbf{APOC1}} & \multirow{2}*{IGSF23} & PLEKHG5, & \multirow{2}*{TBCC} \\
 & & & & &  TNFRSF25 &\\
 & & \textbf{2.38E-14} & \textbf{1.59E-09} & 5.01E-05 & 7.94E-05 & 8.11E-05 \\

 \hline

 & \multicolumn{1}{c}{\multirow{2}*{S1}} & \textbf{APOE} & \textbf{APOC1} & IGSF23 & MTMR2 & HSBP1 \\
 & & \textbf{1.11E-16} & \textbf{2.03E-10} & 3.12E-05 & 9.79E-05 & 1.15E-04 \\  \cline{3-7}
\multicolumn{1}{c}{\multirow{4}{*}{Lin}} & \multicolumn{1}{c}{\multirow{2}*{S2}} & \textbf{APOE} & \textbf{APOC1} & IGSF23 & MTMR2 & GRIP1 \\
 & & \textbf{2.00E-15} & \textbf{8.31E-10} & 6.73E-05 & 1.43E-04 & 1.52E-04 \\ \cline{3-7}
 & \multicolumn{1}{c}{\multirow{2}*{S3}} & \textbf{APOE} & \textbf{APOC1} & IGSF23 & HSBP1 & MTMR2 \\
 & & \textbf{6.33E-15} & \textbf{4.78E-10} & 3.30E-05 & 1.04E-04 & 1.22E-04 \\ \cline{3-7}
 & \multicolumn{1}{c}{\multirow{2}*{S4}} & \textbf{APOE} & \textbf{APOC1} & IGSF23 & MTMR2 & HSBP1 \\
 & & \textbf{4.44E-16} & \textbf{1.99E-10} & 3.18E-05 & 9.92E-05 & 1.13E-04 \\

\hline

& \multicolumn{1}{c}{\multirow{2}{*}{S1}} & \textbf{APOE} & \textbf{APOC1} & IGSF23 & GRIP1 & TBCC \\
 & & \textbf{1.34E-13} & \textbf{5.01E-09} & 4.56E-05 & 1.14E-04 & 1.53E-04 \\ \cline{3-7}
\multicolumn{1}{c}{\multirow{4}{*}{Lap}} & \multicolumn{1}{c}{\multirow{2}*{S2}} & \multirow{2}*{\textbf{APOE}} & \multirow{2}*{\textbf{APOC1}} & \multirow{2}*{IGSF23} & PLEKHG5,& \multirow{2}*{GRIP1} \\
& & & & &  TNFRSF25 &\\
 & & \textbf{2.59E-13} & \textbf{1.23E-08} & 1.31E-04 & 1.40E-04 & 1.65E-04 \\ \cline{3-7}
 & \multicolumn{1}{c}{\multirow{2}*{S3}} & \textbf{APOE} & \textbf{APOC1} & IGSF23 & TBCC & GRIP1 \\
 & & \textbf{2.03E-13} & \textbf{4.80E-09} & 3.88E-05 & 9.40E-05 & 1.20E-04 \\ \cline{3-7}
 & \multicolumn{1}{c}{\multirow{2}*{S4}} & \textbf{APOE} & \textbf{APOC1} & IGSF23 & GRIP1 & TBCC \\
 & & \textbf{1.37E-13} & \textbf{5.95E-09} & 4.58E-05 & 1.15E-04 & 1.53E-04 \\

\hline

& \multicolumn{1}{c}{\multirow{2}{*}{S1}} & \textbf{APOE} & \textbf{APOC1} & EHHADH-AS1 & GRIP1 & GABBR1 \\
 & & \textbf{1.33E-15} & \textbf{2.99E-10} & 1.18E-04 & 1.48E-04 & 1.72E-04 \\ \cline{3-7}
\multicolumn{1}{c}{\multirow{4}{*}{Quad}} & \multicolumn{1}{c}{\multirow{2}*{S2}} & \textbf{APOE} & \textbf{APOC1} & EHHADH-AS1 & GRIP1 & C16orf54 \\
 & & \textbf{1.11E-15} & \textbf{1.67E-09} & 7.80E-05 & 1.32E-04 & 1.96E-04 \\ \cline{3-7}
 & \multicolumn{1}{c}{\multirow{2}*{S3}} & \textbf{APOE} & \textbf{APOC1} & EHHADH-AS1  & GABBR1 & GRIP1 \\
 & & \textbf{4.44E-16} & \textbf{3.55E-10} & 1.38E-04 & 1.44E-04 & 1.45E-04 \\ \cline{3-7}
 & \multicolumn{1}{c}{\multirow{2}*{S4}} & \textbf{APOE} & \textbf{APOC1} & EHHADH-AS1 & GRIP1 & GABBR1 \\
 & & \textbf{4.44E-16} & \textbf{2.23E-10} & 1.16E-04 & 1.48E-04 & 1.69E-04 \\

\hline
\multicolumn{2}{c}{p-value threshold$\dagger$} & 4.46E-07 & 8.91E-07 & 1.34E-06 & 1.78E-06 & 2.23E-06 \\
\hline
\end{tabular}}
\begin{tablenotes}
$\dagger$ FDR-based p-value thresholds were calculated according to \citet{BeHo} under arbitrary dependence assumption, i.e., $\text{Threshold}_i = \frac{\alpha i}{m\sum_{k=1}^m (1/k)}\ (i=1,\ldots,m)$, where $m$ is the number of tests and $\alpha$ is the target FDR.
\end{tablenotes}
\end{table}

We further performed genome-wide gene-based association analyses by using $R^c_{\mbox{het}}$ to consider three types of genetic heterogeneity,  namely genetic heterogeneity due to different sexes, education levels and genetic backgrounds (S2-S4 in Table \ref{realdataanalysis}). When considering heterogeneity across different genetic backgrounds, we randomly selected 200,000 SNPs  from the whole genome to measure the genetic background.  \textit{APOE} and \textit{APOC1} remained to be the only two genome-wide significant genes under the consideration of each type of genetic heterogeneity  (Table \ref{realdataanalysis}).

Although not reaching the genome-wide significance level,  \textit{IGSF23} was most frequently found to be the third ranking gene that is associated with age at AD onset in our analyses (Table \ref{realdataanalysis}). It was also implicated in a recent study of AD genetics \citep{Jansen}.

\section{Discussion}\label{dis}
We have developed a suite of novel genetic association tests for survival outcomes based on the accelerated failure time model. They have correct type I error rates under finite samples of realistic sizes and are able to deal with competing risks and left truncation. The new association tests can account for genetic heterogeneity to improve the power of association discovery.

Several future research directions related to this work are worth pursuing. First, a simple multi-marker gene-gene/gene-environment interaction test can be developed by adding the genetic markers (and environment variable(s) if testing for a gene-environment interaciton) to the covariates of the AFT null model and using the test statistic \eqref{assoc_test_stat} with $\bK$ replaced by a Hadamard product of two kernel matrices corresponding to the main effects of the two marker sets (or the marker set and the environment variable(s) if testing for a gene-environment interaction) respectively. Due to the use of kernels, this test will be more powerful than the Wald test in the AFT model for testing the interaction when the genetic markers are correlated. However, a drawback of this test is that it assumes a parametric form of the main effects of the two marker sets (or the marker set and the environment variable(s)) to adapt the derivation of the asymptotic null distributions for the association tests to the interaction test. This assumption might not be true in practice. It would be ideal to develop an interaction test that  nonparametrically models the main effects. Second, genes in a biological pathway are connected due to regulatory interactions. By leveraging this network (graph) information, which can be obtained from pathway databases like KEGG \citep{KEGG}, we can increase the power of our tests for association  involving gene sets. Third, our tests were developed for data from unrelated subjects. It is worthwhile to extend them to related individuals, which inevitably exist in family studies and biobank data. Such an extension might be accomplished by incorporating a frailty term that captures relatedness into the AFT kernel machine regression \citep{SiCa2018}.



\makeatletter
\renewcommand\thesection{}
\renewcommand\thesubsection{\@Alph\c@section.\@arabic\c@subsection}
\makeatother

\setcounter{section}{0}
\section{
Appendix A: Derivation of the Large-Sample Null Distribution of $R$}

Write $\bE_1$ as $\bE_1=(\bE_{11},\ldots,\bE_{1n})$. Then
\be
\bE_1\widetilde{\bM}&=&\sum_{i=1}^n\bE_{1i}\widetilde{M}_i \nonumber \\
&=&\sum_{i=1}^n\bE_{1i}\int_{-\infty}^\infty\left\{dN_i(\widehat{\bbeta},t)-\frac{\nu_i(\widehat{\bbeta},t)Y_i(\widehat{\bbeta},t)}{\sum_{j=1}^n\nu_j(\widehat{\bbeta},t)Y_j(\widehat{\bbeta},t)}\sum_{j=1}^ndN_j(\widehat{\bbeta},t)\right\}\nonumber \\
&=&\sum_{i=1}^n\int_{-\infty}^\infty\left\{\bE_{1i}-\frac{\sum_{j=1}^n\bE_{1j}\nu_j(\widehat{\bbeta},t)Y_j(\widehat{\bbeta},t)}{\sum_{j=1}^n\nu_j(\widehat{\bbeta},t)Y_j(\widehat{\bbeta},t)}\right\}dN_i(\widehat{\bbeta},t) \nonumber \\
&\equiv&\bQ_n(\widehat{\bbeta}).
\ee
Under the null, $\widehat{\bbeta}$ is $n^{1/2}$-consistent for $\bbeta$ \citep{LaYi}. Using similar arguments to the proof of Theorem 1(ii) in \citet{LaYi}, it can be shown that
\bq\label{asymplin}
n^{-1/2}\bE_1(\widetilde{\bM}-\widehat{\bM})=n^{-1/2}\{\bQ_n(\widehat{\bbeta})-\bQ_n(\bbeta)\}=\bB n^{1/2}(\widehat{\bbeta}-\bbeta)+o_p(1+n^{1/2}\|\widehat{\bbeta}-\bbeta\|),
\eq
where $\bB$ is the asymptotic slope matrix of $n^{-1}\bQ_n(\bbeta)$. Following \citet{ZeLi2008}, we use a least squares method to estimate $\bB$. Specifically, let $\widetilde{\bbeta}=\widehat{\beta}+n^{-1/2}\bW$, where $\bW\sim N(\bzero,\bI_q)$ and $\bI_q$ is a $q \times q$ identity matrix. Equation \eqref{asymplin} implies that
\bq
n^{-1/2}\{\bQ_n(\widetilde{\bbeta})-\bQ_n(\widehat{\bbeta})\}=\bB n^{1/2}(\widetilde{\bbeta}-\widehat{\bbeta})+o_p(1)=\bB\bW+o_p(1).
\eq
The least squares method has the following steps:
\begin{itemize}
    \item[\textbf{Step 1:}] Generate $L$, say 10,000, independent realizations of $\bW$, denoted by $\bW_1,\ldots,\bW_L$.
\item[\textbf{Step 2:}] Calculate $n^{-1/2}\{\bQ_n(\widehat{\bbeta}+n^{-1/2}\bW_l)-\bQ_n(\widehat{\bbeta})\}$ $(l=1,\ldots,L).$
\item[\textbf{Step 3:}] For $j=1,\ldots,m$, regress $n^{-1/2}\{\bQ_n(\widehat{\bbeta}+n^{-1/2}\bW_l)-\bQ_n(\widehat{\bbeta})\}_j$ onto $\bW_l$ $(l=1,\ldots,L)$ to estimate $\bB_j$, the $j$-th row of $\bB$, using the least squares estimation.
\end{itemize}
Denote the estimator of $\bB$ by $\widehat{\bB}$. Recall that $\widehat{\bbeta}$ is obtained from Eq. (5) in \citet{ChXu} with log-rank weights,
\bq
\bU_n(\bbeta)\equiv\frac{1}{n}\sum_{i=1}^n\int_{-\infty}^{\infty}\left\{\bZ_i-\frac{\sum_{j=1}^n\bZ_j\nu_j(\bbeta,t)Y_j(\bbeta,t)}{\sum_{j=1}^n\nu_j(\bbeta,t)Y_j(\bbeta,t)}\right\}dN_i(\bbeta,t)=\bzero.
\eq
\citet{ChXu} showed that
\bq \label{beta_asymp_expr}
n^{1/2}(\widehat{\bbeta}-\bbeta)=-\bA^{-1}n^{1/2}\bU_n(\bbeta)+o_p(1),
\eq
where $\bA$ is the asymptotic slope matrix of $\bU_n(\bbeta)$. $\bA$ can be estimated using a similar method to that for estimating $\bB$. Specifically, we carry out the following steps:
\begin{itemize}
    \item[\textbf{Step I:}] Generate $\widetilde{L}$, say 10,000, independent realizations of $\bS\sim N(\bzero,\bI_q)$, denoted by $\bS_1,\ldots,\bS_{\widetilde{L}}$.
\item[\textbf{Step II:}] Calculate $n^{1/2}\bU_n(\widehat{\bbeta}+n^{-1/2}\bS_l)$ $(l=1,\ldots,\widetilde{L}).$
\item[\textbf{Step III:}] For $j=1,\ldots,q$, regress $n^{1/2}U_{nj}(\widehat{\bbeta}+n^{-1/2}\bS_l)$ onto $\bS_l$ $(l=1,\ldots,\widetilde{L})$ to estimate $\bA_j$, the $j$-th row of $\bA$, using the least squares estimation.
\end{itemize}
Denote the estimator of $\bA$ by $\widehat{\bA}$. Combining \eqref{asymplin} and \eqref{beta_asymp_expr}, we have
\bq\label{EMtilde-EMhat}
\bE_1(\widetilde{\bM}-\widehat{\bM})=-\bB\bA^{-1}nU_n(\bbeta)+o_p(n^{1/2}).
\eq

Let $M_i(t)=N_i(\bbeta,t)-\int_{-\infty}^t\nu_i(\bbeta,u)Y_i(\bbeta,u)d\Lambda_\varepsilon(u)$, which is a martingale under the null. It is easy to show that
\be
\bU_n(\bbeta)&=&\frac{1}{n}\sum_{i=1}^n\int_{-\infty}^{\infty}\left\{\bZ_i-\frac{\sum_{j=1}^n\bZ_j\nu_j(\bbeta,t)Y_j(\bbeta,t)}{\sum_{j=1}^n\nu_j(\bbeta,t)Y_j(\bbeta,t)}\right\}dM_i(\bbeta,t) \nonumber \\
&=&\frac{1}{n}\bZ^T\int_{-\infty}^{\infty}(\bI_n-\bV(\bbeta,t)\bdone^T)d\bM(t), \label{U_martingale}
\ee
where $\bdone$ is a $n$-dimension vector of 1's, $\bZ=(\bZ_1,\ldots,\bZ_n)^T$,\\ $\bV(\bbeta,t)=(\nu_1(\bbeta,t)Y_1(\bbeta,t),\ldots,\nu_1(\bbeta,t)Y_n(\bbeta,t))^T\{\sum_{j=1}^n\nu_j(\bbeta,t)Y_j(\bbeta,t)\}^{-1}$, and  $\bM(t)=(M_1(t),\ldots,M_n(t))^T$. It is also easy to show that
\bq\label{Mhat_martingale}
\widehat{\bM}=\int_{-\infty}^{\infty}(\bI_n-\bV(\bbeta,t)\bdone^T)d\bM(t).
\eq
Combining \eqref{EMtilde-EMhat}, \eqref{U_martingale} and \eqref{Mhat_martingale} , we have
\be
\bE_1\widetilde{\bM}&=&\bE_1\widehat{\bM}+\bE_1(\widetilde{\bM}-\widehat{\bM}) \nonumber \\
&=&\int_{-\infty}^{\infty}(\bE_1-\bB\bA^{-1}\bZ^T)(\bI_n-\bV(\bbeta,t)\bdone^T)d\bM(t)+o_p(n^{1/2})
\ee

Since $\bM(t)$ is a vector of independent martingales under the null, whose compensators are $\bLambda(t)=\int_{-\infty}^td\bLambda(t)\equiv(\int_{-\infty}^td\Lambda_1(t),\ldots,\int_{-\infty}^td\Lambda_n(t))^T$, where $d\Lambda_i(t)=\nu_i(\bbeta,t)Y_i(\bbeta,t)d\Lambda_\varepsilon(t)$ $(i=1,\ldots,n)$, we have by the martingale theory \citep[Chapters 2 and 5]{FlHa} that
\bq
Cov(\bE_1\widetilde{\bM})\approx\int_{-\infty}^{\infty}(\bE_1-\bB\bA^{-1}\bZ^T)(\bI_n-\bV(\bbeta,t)\bdone^T)\mbox{diag}(d\bLambda(t))(\bI_n-\bdone \bV(\bbeta,t)^T)(\bE_1^T-ZA^{-T}\bB^T),
\eq
where $\mbox{diag}(\ba)$ represents a diagonal matrix with $\ba$ as the diagonal, and that $\bE_1\widetilde{\bM}$ is approxmiately multivariate normal with mean zero. An estimator of $Cov(\bE_1\widetilde{\bM})$ is
\bq
\widehat{Cov}(\bE_1\widetilde{\bM})=\int_{-\infty}^{\infty}(\bE_1-\widehat{\bB}\widehat{\bA}^{-1}\bZ^T)(\bI_n-\bV(\widehat{\bbeta},t)\bdone^T)\mbox{diag}(d\widetilde{\bLambda}(t))(\bI_n-\bdone \bV(\widehat{\bbeta},t)^T)(\bE_1^T-\bZ\widehat{\bA}^{-T}\widehat{\bB}^T),
\eq
where $d\widetilde{\bLambda}(t)=(d\widetilde{\Lambda}_1(t),\ldots,d\widetilde{\Lambda}_n(t))^T$ and $d\widetilde{\Lambda}_i(t)=\nu_i(\widehat{\bbeta},t)Y_i(\widehat{\bbeta},t)\sum_{j=1}^ndN_j(\widehat{\bbeta},t)\{\sum_{j=1}^n\nu_j(\widehat{\bbeta},t)Y_j(\widehat{\bbeta},t)\}^{-1}$ $(i=1,\ldots,n)$. By algebra, we can simplify $\widehat{Cov}(\bE_1\widetilde{\bM})$ as
\bq
\widehat{Cov}(\bE_1\widetilde{\bM})=\int_{-\infty}^{\infty}(\bE_1-\widehat{\bB}\widehat{\bA}^{-1}\bZ^T)\{\mbox{diag}(d\widetilde{\bLambda}(t))-\bV(\widehat{\bbeta},t)\bdone^T d\bN(\widehat{\bbeta},t) \bV(\widehat{\bbeta},t)^T\}(\bE_1^T-\bZ\widehat{\bA}^{-T}\widehat{\bB}^T),
\eq
where $\bN(\widehat{\bbeta},t)=(N_1(\widehat{\bbeta},t),\ldots,N_n(\widehat{\bbeta},t))^T$. Take an eigendecomposition of $\widehat{Cov}(\bE_1\widetilde{\bM})$,
\bq\widehat{Cov}(\bE_1\widetilde{\bM})=\Gamma\begin{bmatrix}
    \lambda_{1} & & \\
    & \ddots & \\
    & & \lambda_{m}
  \end{bmatrix}\Gamma^T,\eq
  where $\Gamma$ is a $m\times m$ orthogonal matrix. Together with the asymptotic normality of $\bE_1\widetilde{\bM}$ , we have
  \bq\bE_1\widetilde{\bM}\stackrel{\text{d}}{\approx}\Gamma\begin{bmatrix}
    \lambda_{1}^{1/2} & & \\
    & \ddots & \\
    & & \lambda_{m}^{1/2}
  \end{bmatrix}\bchi,  \eq
  where $\bchi\equiv(\chi_{11},\ldots,\chi_{1m})^T\sim N(\bzero,\bI_m)$.
  Thus,  \bq R=\widetilde{\bM}^T\bK\widetilde{\bM}=\widetilde{\bM}^T\bE_1^T\bE_1\widetilde{\bM}\stackrel{\text{d}}{\approx}\sum_{j=1}^m\lambda_j\chi_{1j}^2,\eq
where $\chi_{1j}^2$'s are independent chi-square variables with degree 1.

\section{Appendix B: Additional Simulations}
In the following additional simulations except the ones in Section \ref{sec_coxKM}, a two-step procedure was used to simulate SNPs as the genetic markers under testing: 1) sample $n$ vectors independently from a multivariate normal distribution with a zero mean and the covariance matrix being $\bSigma_{p\times p} = \{0.5^{|k-l|}\}$; 2) categorize each component of every multivariate normal vector into three levels labeled with 0, 1 and 2 using the cut-off values that were selected to satisfy the Hardy-Weinberg equilibrium (HWE) and the minor allele frequency (MAF) simulated from $Beta(2,5)$. In the simulations of Section \ref{sec_coxKM}, we sampled SNPs from the genotype data of the 1000 Genomes Project (phase 3) \citep{1000G} as the genetic markers.

\subsection{Empirical size and power of  $R_{\mbox{het}}$ in the presence of genetic heterogeneity across two latent sub-populations}
\label{3.2.2}

In this simulation, we investigated the empirical size and power of $R_{\mbox{het}}$ as well as $R$ under genetic heterogeneity across two latent sub-populations with equal proportions. The survival time of the $j$-th subject in the $i$-th sub-population was generated from the following AFT models for the cause-specific hazard functions,

\begin{equation}
    \lambda_1 (t|\bZ_{ij},\bG_{ij}) = \lambda_0(t\cdot \exp \{-\sum_{k=1}^p G_{ijk}\beta_{ik} - 0.1 Z_{ij1} - 0.1 Z_{ij2}\}) \exp \{-\sum_{k=1}^p G_{ijk}\beta_{ik} - 0.1 Z_{ij1} - 0.1 Z_{ij2}\},
\label{aftmod31}
\end{equation}

\begin{equation}
    \lambda_2 (t|\bZ_{ij},\bG_{ij}) = \lambda_0(t\cdot \exp \{-\sum_{k=1}^p 0.02 G_{ijk} - 0.2 Z_{ij1} - 0.2 Z_{ij2}\}) \exp \{-\sum_{k=1}^p 0.02 G_{ijk} - 0.2 Z_{ij1} - 0.2 Z_{ij2}\},
\label{aftmod32}
\end{equation}
where $\beta_{i1}=\ldots =\beta_{ip}\ (i=1,2)$ represent the effects of $G_k$'s $(k=1,\ldots,p)$ in Sub-population $i$ and vary depending on the heterogeneity scenario. A continuous variable, $X_{ij}$ $(i=1,2)$, was simulated to infer the sub-population. Specifically, $X_{ij}=I(i=1)+1+e_{ij}$, where $e_{ij} \sim N(0,0.5)$. We set $\beta_{ik}=0$ $(i=1,2;\ k=1,\ldots,p)$ in the size assessment, while different values were assigned to $\beta_{ik}$'s to represent different heterogeneity scenarios in the power evaluation, as shown in Table \ref{hwv_2sub_power}. The IBS kernel was used to measure the genetic similarity in $R_{\mbox{het}}$ and $R$, and the Gaussian kernel was applied to $X_{ij}$ to measure the sub-population similarity in $R_{\mbox{het}}$. Table \ref{hwv_2sub_t1e} shows that the empirical sizes of both $R_{\mbox{het}}$ and  $R$ are close to the nominal level. Table \ref{hwv_2sub_power} shows that the power of $R_{\mbox{het}}$ increases with the sample size and the heterogeneity size, measured by $|\beta_{1k}-\beta_{2k}|$, the genetic effect difference between the two sub-populations. Also, Table \ref{hwv_2sub_power} indicates that when there is no genetic heterogeneity  (Scenario T1) , $R_{\mbox{het}}$ has a smaller power than $R$. However, $R_{\mbox{het}}$ is more powerful when there exists genetic heterogeneity between the two latent sub-populations (Scenarios T2 - T4).

\begin{table}[H]
\setlength\tabcolsep{2pt}
\caption{Empirical sizes of $R$ and $R_{\mbox{het}}$ in testing genetic effects under genetic heterogeneity across two latent sub-populations, covariate adjustment and left truncation.}
\label{hwv_2sub_t1e}
\begin{tabular*}{\textwidth}{@{\extracolsep{\fill}} l *{3}{d{2.4}} }
\hline
& \multicolumn{2}{c}{Empirical Size} \\ \cline{2-3}
 & \multicolumn{1}{c}{p=3, n=400} & \multicolumn{1}{c}{p=3, n=500} \\ \cline{2-3}
\multicolumn{1}{l}{$R_{\mbox{het}}$} & \multicolumn{1}{c}{0.047} & \multicolumn{1}{c}{0.050} \\
\multicolumn{1}{l}{$R$} & \multicolumn{1}{c}{0.052} & \multicolumn{1}{c}{0.058} \\ \cline{2-3}
& \multicolumn{1}{c}{p=5, n=400} & \multicolumn{1}{c}{p=5, n=500} \\ \cline{2-3}
\multicolumn{1}{l}{$R_{\mbox{het}}$} & \multicolumn{1}{c}{0.053} & \multicolumn{1}{c}{0.044} \\
\multicolumn{1}{l}{$R$} & \multicolumn{1}{c}{0.045} & \multicolumn{1}{c}{0.049} \\
\hline
\end{tabular*}
\end{table}

\begin{table}[H]
\setlength{\tabcolsep}{6pt}
\renewcommand{\arraystretch}{0.8}
\caption{Powers of $R$ and $R_{\mbox{het}}$ in testing genetic effects under genetic heterogeneity across two latent sub-populations, covariate adjustment and left truncation. Various heterogeneity scenarios were considered, determined by the values of $\beta_{1k}$ and $\beta_{2k}$, including the same effect size and the same effect direction (T1), identical sizes but opposite directions (T2), no effect in one sub-population while positive effect in the other (T3), and different sizes but the same direction (T4).}
\label{hwv_2sub_power}
{\begin{tabular}{lccccccccc}
\hline
 & & \multicolumn{8}{c}{Heterogeneity Scenario} \\ \cline{3-10}
 & & \multicolumn{2}{c}{T1} & \multicolumn{2}{c}{T2} & \multicolumn{2}{c}{T3} & \multicolumn{2}{c}{T4} \\ \cline{3-10}
& \multicolumn{1}{c}{$\beta_{1k}$} & 0.04 & 0.08 & -0.05 & -0.1 & 0 & 0       & 0.03 & 0.03 \\
& \multicolumn{1}{c}{$\beta_{2k}$} & 0.04 & 0.08 & 0.05 & 0.1   & 0.08 & 0.12 & 0.08 & 0.1 \\ \cline{1-10}
\multirow{2}{*}{p=3,n=400}  & \multicolumn{1}{l}{$R_{\mbox{het}}$} & 0.252 & 0.768 & 0.775 & 0.972 & 0.719 & 0.906 & 0.601 & 0.784 \\
                            & \multicolumn{1}{l}{$R$} & 0.376 & 0.891 & 0.056 & 0.043 & 0.370 & 0.636 & 0.596 & 0.746 \\ \cline{1-10}
\multirow{2}{*}{p=3,n=500}  & \multicolumn{1}{l}{$R_{\mbox{het}}$} & 0.327 & 0.825 & 0.826 & 0.979 & 0.768 & 0.945 & 0.703 & 0.868 \\
                            & \multicolumn{1}{l}{$R$} & 0.376 & 0.927 & 0.057 & 0.047 & 0.467 & 0.716 & 0.702 & 0.817 \\ \cline{1-10}
\multirow{2}{*}{p=5,n=400}  & \multicolumn{1}{l}{$R_{\mbox{het}}$} & 0.403 & 0.900 & 0.984 & 1.000 & 0.950 & 0.994 & 0.855 & 0.958 \\
                            & \multicolumn{1}{l}{$R$} & 0.594 & 0.976 & 0.062 & 0.069 & 0.557 & 0.775 & 0.829 & 0.894 \\ \cline{1-10}
\multirow{2}{*}{p=5,n=500}  & \multicolumn{1}{l}{$R_{\mbox{het}}$} & 0.467 & 0.968 & 0.988 & 1.000 & 0.968 & 0.998 & 0.935 & 0.979 \\
                            & \multicolumn{1}{l}{$R$} & 0.729 & 0.995 & 0.047 & 0.068 & 0.659 & 0.866 & 0.905 & 0.958 \\ \hline
\end{tabular}}
\end{table}

\subsection{Empirical size and power of  $R_{\mbox{het}}$ in the presence of genetic heterogeneity across twenty latent sub-populations}
\label{3.2.3}

In this simulation, we investigated the performances of the tests $R_{\mbox{het}}$ and $R$ under similar settings to the previous section but increased the number of latent sub-populations to 20 with equal proportions. The  effects of $G_k$'s $(k=1,\ldots,p)$ in each of the 20 sub-populations were set to satisfy $\beta_{i1}=\ldots =\beta_{ip}$ $(i=1,\ldots,20)$. $\beta_{i1}$'s $(i=1,\ldots,20)$ were zero in the empirical size assessment and were sampled from a uniform distribution with mean $\mu_{\beta}$ and variance $\sigma_{\beta}^2$ in the power assessment. We simulated 25 covariates, $X_{d}$'s $(d=1,\ldots,25)$, to infer the sub-population. The $X_d$ $(d=1,\ldots,25)$ of the $j$-th subject in $i$-th sub-population was generated by $X_{ijd}=a_{id}+\delta_{ij}$, where $a_{id}$'s $(i=1,\ldots,20)$ were randomly sampled from $\{1,\ldots,20\}$ without replacement and $\delta_{ij} \sim N(0,0.5)$. We used the IBS kernel to measure the genetic similarity in $R_{\mbox{het}}$ and $R$ and the Gaussian kernel to measure the sub-population similarity in $R_{\mbox{het}}$. Table \ref{hwv_20sub_t1e} shows that the empirical sizes of  $R$ and $R_{\mbox{het}}$ are close to the nominal level. Table \ref{hwv_20sub_power} shows that $R_{\mbox{het}}$ is more powerful than $R$ in the presence of genetic heterogeneity across twenty latent sub-populations, and the power advantage gets more obvious as the genetic heterogeneity size ($\sigma_{\beta}$) increases.

\begin{table}[H]
\setlength\tabcolsep{2pt}
\caption{Empirical sizes of $R$ and $R_{\mbox{het}}$ in testing genetic effects under genetic heterogeneity across twenty latent sub-populations, covariate adjustment and left truncation.}
\label{hwv_20sub_t1e}
\begin{tabular*}{\textwidth}{@{\extracolsep{\fill}} l *{3}{d{2.4}} }
\hline
 & \multicolumn{2}{c}{Empirical Size} \\ \cline{2-3}
 & \multicolumn{1}{c}{p=3, n=400} & \multicolumn{1}{c}{p=3, n=500} \\ \cline{2-3}
 \multicolumn{1}{l}{$R_{\mbox{het}}$} & \multicolumn{1}{c}{0.045} & \multicolumn{1}{c}{0.043} \\
 \multicolumn{1}{l}{$R$} & \multicolumn{1}{c}{0.046} & \multicolumn{1}{c}{0.042} \\ \cline{2-3}
 & \multicolumn{1}{c}{p=5, n=400} & \multicolumn{1}{c}{p=5, n=500} \\ \cline{2-3}
\multicolumn{1}{l}{$R_{\mbox{het}}$} & \multicolumn{1}{c}{0.043} & \multicolumn{1}{c}{0.044} \\
\multicolumn{1}{l}{$R$} & \multicolumn{1}{c}{0.042} & \multicolumn{1}{c}{0.048} \\
\hline
\end{tabular*}
\end{table}

\begin{table}[H]
\setlength\tabcolsep{2pt}
\caption{Powers of $R_{\mbox{het}}$ and $R$ under genetic heterogeneity across twenty latent sub-populations, covariate adjustment and left truncation.}
\label{hwv_20sub_power}
\begin{tabular*}{\textwidth}{@{\extracolsep{\fill}} l *{5}{d{2.4}} }
\hline
 & \multicolumn{4}{c}{Power} \\ \cline{2-5}
 & \multicolumn{2}{c}{p=3, n=400} & \multicolumn{2}{c}{p=3, n=500} \\ \cline{2-5}
\multicolumn{1}{c}{($\mu_{\beta}$, $\sigma_{\beta}$)} & \multicolumn{1}{c}{(0.02, 0.04)} & \multicolumn{1}{c}{(0.02, 0.08)} & \multicolumn{1}{c}{(0.02, 0.04)} & \multicolumn{1}{c}{(0.02, 0.08)} \\ \cline{2-5}
\multicolumn{1}{l}{$R_{\mbox{het}}$} & \multicolumn{1}{c}{0.396} & \multicolumn{1}{c}{0.742} & \multicolumn{1}{c}{0.448} & \multicolumn{1}{c}{0.843} \\
\multicolumn{1}{l}{$R$} & \multicolumn{1}{c}{0.336} & \multicolumn{1}{c}{0.506} & \multicolumn{1}{c}{0.390} & \multicolumn{1}{c}{0.616} \\ \cline{2-5}
 & \multicolumn{2}{c}{p=5, n=400} & \multicolumn{2}{c}{p=5, n=500} \\ \cline{2-5}
\multicolumn{1}{c}{($\mu_{\beta}$, $\sigma_{\beta}$)} & \multicolumn{1}{c}{(0.02, 0.04)} & \multicolumn{1}{c}{(0.02, 0.08)} & \multicolumn{1}{c}{(0.02, 0.04)} & \multicolumn{1}{c}{(0.02, 0.08)} \\ \cline{2-5}
\multicolumn{1}{l}{$R_{\mbox{het}}$} & \multicolumn{1}{c}{0.679} & \multicolumn{1}{c}{0.955} & \multicolumn{1}{c}{0.785} & \multicolumn{1}{c}{0.988} \\
\multicolumn{1}{l}{$R$} & \multicolumn{1}{c}{0.473} & \multicolumn{1}{c}{0.634} & \multicolumn{1}{c}{0.581} & \multicolumn{1}{c}{0.754} \\ \hline
\end{tabular*}
\end{table}

\subsection{Empirical size and power of  $R_{\mbox{het}}$ in the presence of genetic heterogeneity across individual genome profiles}
\label{3.2.4}

In this simulation, we investigated the performances of $R_{\mbox{het}}$ and $R$ when the sub-population structure is "continuous". Specifically, we let the genetic effect vary across individual genome profiles instead of a small number of sub-populations, e.g., males and females. The survival time of the $i$-th $(i=1,\ldots,n)$ subject was generated from the models \eqref{aftmod31} and \eqref{aftmod32}, where $\beta_{i1}=\beta_{i2}=\ldots=\beta_{ip}=0$ in the size assessment and were randomly sampled from a uniform distribution with mean $\mu_{\beta}$ and variance $\sigma^2_{\beta}$ in the power assessment. The values of $\mu_{\beta}$ and $\sigma^2_{\beta}$ vary  in different simulation scenarios. We simulated a set of 1000 SNPs for each subject, $\{X_{id}\}_{d=1}^{1000}$ $(i=1,\ldots,n)$, to represent the genome profile. For each $1\le d\le 1000$, $\bX_d=(X_{1d},\ldots,X_{nd})^T$ was generated in two steps: 1) sample $\Tilde{\bX}_d=(\widetilde{X}_{1d},\ldots,\widetilde{X}_{nd})^T$ from a multivariate normal distribution, $MVN(\bzero,\bSigma)$, where $\bSigma$ is an $n\times n$ covariance matrix with the $(i,j)$-th element being $\Sigma_{ij}=I(i=j)$ under the null hypothesis (i.e., no genetic association) and  $\Sigma_{ij}=\exp(-|\beta_{i1}-\beta_{j1}|/\sigma_{\beta})$ under the alternative; 2) $\bX_{d}$  is then obtained by categorizing each element of $\widetilde{\bX}_{d}$ into three levels, 0, 1 and 2, using rank-based cut-off values selected to achieve the Hardy-Weinberg equilibrium and a pre-specified minor allele frequency that was randomly sampled from $Beta(1,3)$. We used the IBS kernel to measure both the genetic and the sub-population similarities in $R_{\mbox{het}}$ and $R$. Table \ref{hwv_genome_t1e} shows that the empirical sizes of both tests are around the nominal level. Table \ref{hwv_genome_power} shows that $R_{\mbox{het}}$ is more powerful than $R$ under genetic heterogeneity across individual genome profiles, and the power advantage increases with the genetic heterogeneity size ($\sigma_{\beta}$).

\begin{table}[H]
\setlength\tabcolsep{2pt}
\caption{Empirical sizes of $R_{\mbox{het}}$ and $r$ in testing genetic effects under genetic heterogeneity across individual genome profiles, covariate adjustment and left truncation.}
\label{hwv_genome_t1e}
\begin{tabular*}{\textwidth}{@{\extracolsep{\fill}} l *{3}{d{2.4}} }
\hline
 & \multicolumn{2}{c}{Empirical Size} \\ \cline{2-3}
 & \multicolumn{1}{c}{p=3, n=400} & \multicolumn{1}{c}{p=3, n=500} \\ \cline{2-3}
\multicolumn{1}{l}{$R_{\mbox{het}}$} & \multicolumn{1}{c}{0.052} & \multicolumn{1}{c}{0.058} \\
\multicolumn{1}{l}{$R$} & \multicolumn{1}{c}{0.053} & \multicolumn{1}{c}{0.058} \\ \cline{2-3}
 & \multicolumn{1}{c}{p=5, n=400} & \multicolumn{1}{c}{p=5, n=500} \\ \cline{2-3}
 \multicolumn{1}{l}{$R_{\mbox{het}}$} & \multicolumn{1}{c}{0.046} & \multicolumn{1}{c}{0.049} \\
  \multicolumn{1}{l}{$R$} & \multicolumn{1}{c}{0.047} & \multicolumn{1}{c}{0.051} \\
\hline
\end{tabular*}
\end{table}

\begin{table}[H]
\setlength\tabcolsep{2pt}
\caption{Powers of $R_{\mbox{het}}$ and $R$ under genetic heterogeneity across individual genome profiles, covariate adjustment and left truncation.}
\label{hwv_genome_power}
\begin{tabular*}{\textwidth}{@{\extracolsep{\fill}} l *{5}{d{2.4}} }
\hline
 & \multicolumn{4}{c}{Power} \\ \cline{2-5}
 & \multicolumn{2}{c}{p=3, n=400} & \multicolumn{2}{c}{p=3, n=500} \\ \cline{2-5}
\multicolumn{1}{c}{($\mu_{\beta}$, $\sigma_{\beta}$)} & \multicolumn{1}{c}{(0.03, 0.02)} & \multicolumn{1}{c}{(0.03, 0.04)} & \multicolumn{1}{c}{(0.03, 0.02)} & \multicolumn{1}{c}{(0.03, 0.04)} \\ \cline{2-5}
\multicolumn{1}{l}{$R_{\mbox{het}}$} & \multicolumn{1}{c}{0.253} & \multicolumn{1}{c}{0.328} & \multicolumn{1}{c}{0.308} & \multicolumn{1}{c}{0.400} \\
\multicolumn{1}{l}{$R$} & \multicolumn{1}{c}{0.241} & \multicolumn{1}{c}{0.242} & \multicolumn{1}{c}{0.290} & \multicolumn{1}{c}{0.292} \\ \cline{2-5}
 & \multicolumn{2}{c}{p=5, n=400} & \multicolumn{2}{c}{p=5, n=500} \\ \cline{2-5}
\multicolumn{1}{c}{($\mu_{\beta}$, $\sigma_{\beta}$)} & \multicolumn{1}{c}{(0.03, 0.02)} & \multicolumn{1}{c}{(0.03, 0.04)} & \multicolumn{1}{c}{(0.03, 0.02)} & \multicolumn{1}{c}{(0.03, 0.04)} \\ \cline{2-5}
\multicolumn{1}{l}{$R_{\mbox{het}}$} & \multicolumn{1}{c}{0.449} & \multicolumn{1}{c}{0.649} & \multicolumn{1}{c}{0.550} & \multicolumn{1}{c}{0.756} \\
\multicolumn{1}{l}{$R$} & \multicolumn{1}{c}{0.390} & \multicolumn{1}{c}{0.381} & \multicolumn{1}{c}{0.480} & \multicolumn{1}{c}{0.477} \\ \hline
\end{tabular*}
\end{table}

\subsection{Performance of $R$ under model misspecification}
\begin{figure}[H]
\captionsetup[subfigure]{justification=centering}
\centering
\subcaptionbox{$n=400, p=5$}{\includegraphics[width=0.5\textwidth]{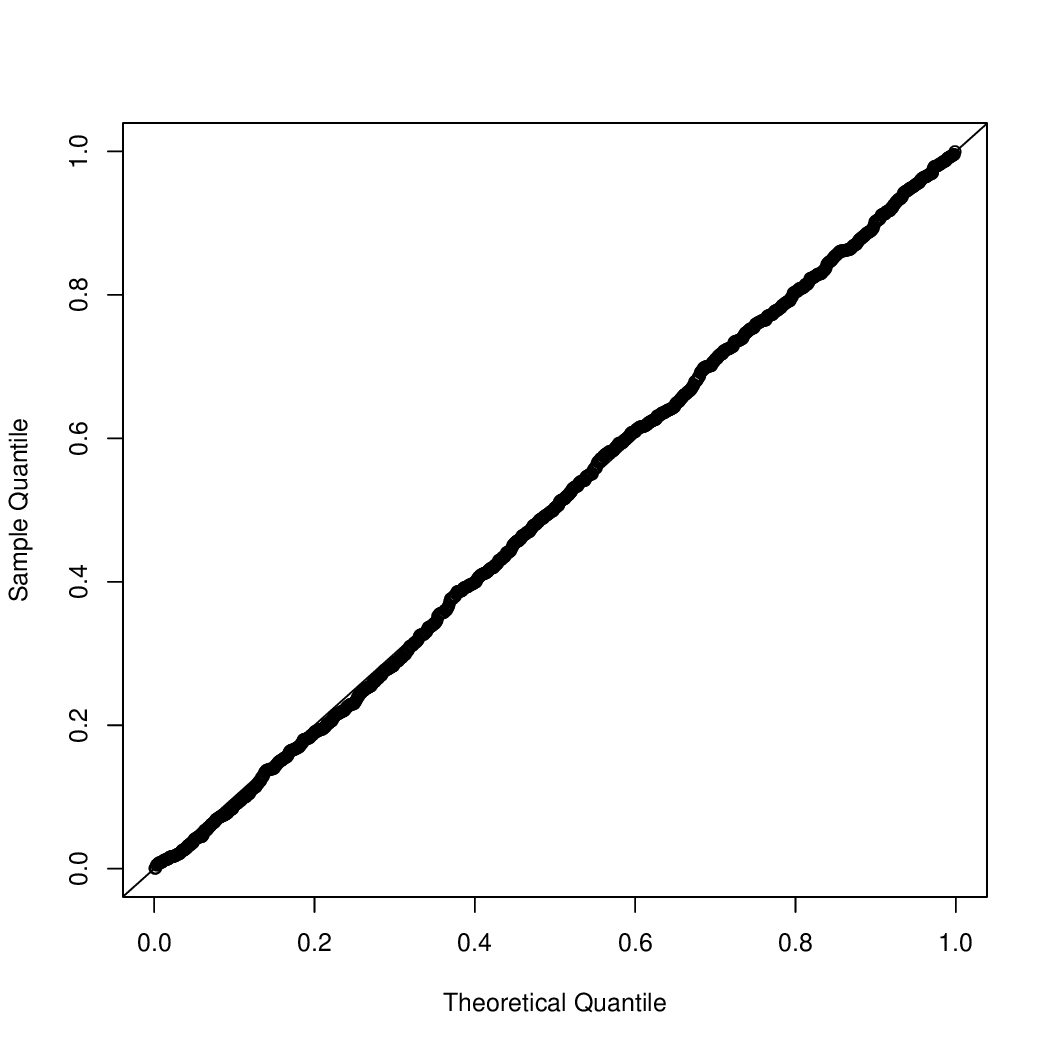}}%
\hfill 
\subcaptionbox{$n=500, p=5$}{\includegraphics[width=0.5\textwidth]{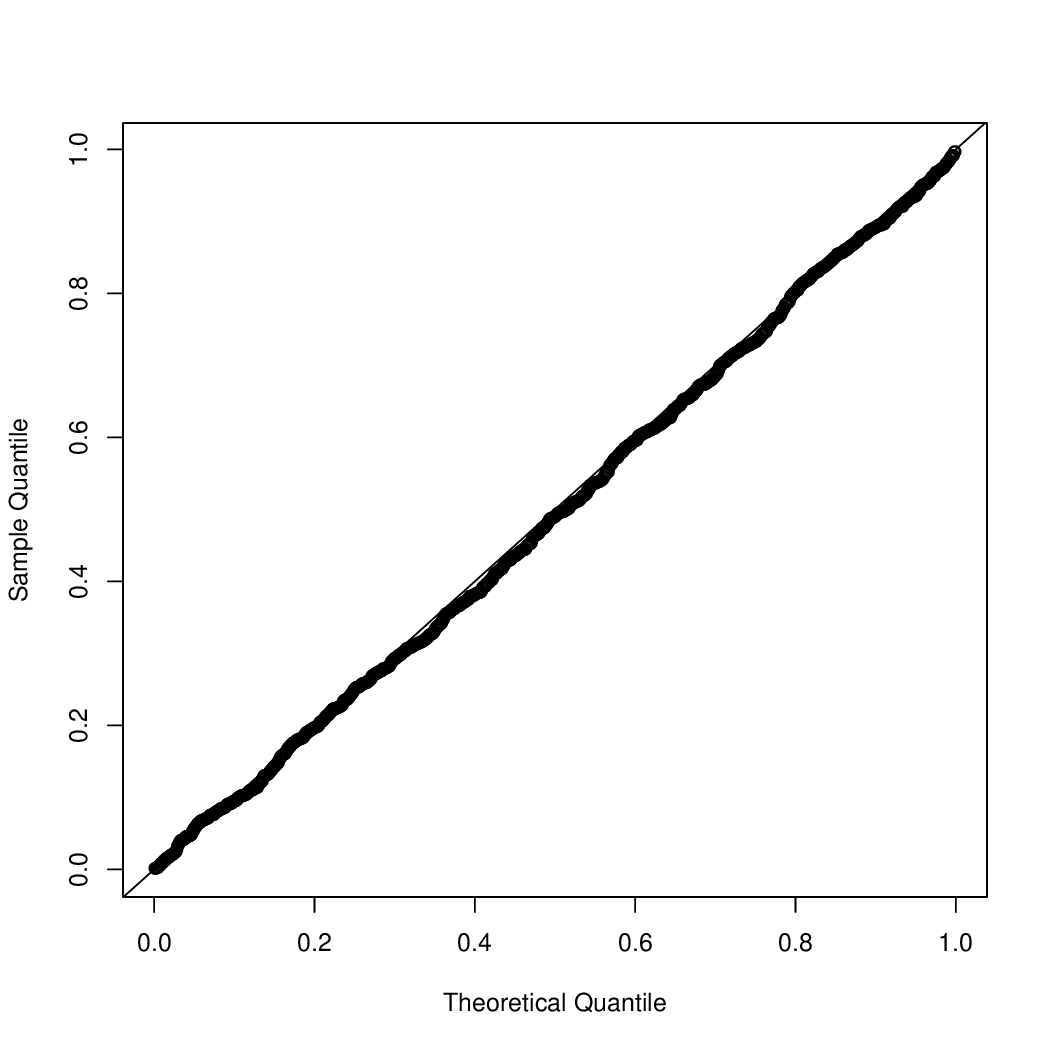}}%
\caption{Uniform Q-Q plots of the null p-values of $R$, with data generated from the Cox model.}
\label{R_robustness_figure}
\end{figure}

In this simulation, we assessed the robustness of $R$ against model misspecification. The competing risks data were generated from the following Cox models for the cause-specific hazard functions,
\begin{equation}
    \lambda_1 (t|\bZ,\bG) = 0.5 (t + t^2) \exp \{\sum_{j=1}^p \beta_j G_j + \sum_{k=1}^2 0.05 Z_k\}
\label{aftmod_robust1}
\end{equation}
and
\begin{equation}
    \lambda_2 (t|\bZ,\bG) = 0.1 (t + t^2) \exp \{\sum_{j=1}^p 0.2 G_j + \sum_{k=1}^2 0.15 Z_k\},
\label{aftmod_robust2}
\end{equation}
where $\beta_j = 0.1$ and $0$ $(j=1,\ldots,p)$ in the power evaluation and size assessment, respectively. The censoring time, left truncation time and adjustment covariates were generated in the same way as in the simulation of association testing without considering genetic heterogeneity.
Table \ref{wv_robustness_table} shows that the empirical size of $R$ is close to the nominal level under various $n$'s and $p$'s and the power of $R$ increases with the sample size. Figure \ref{R_robustness_figure} shows that the null distribution of $R$'s p-value is still close to $U[0,1]$.

\begin{table}[H]
\centering
\setlength\tabcolsep{36pt}
\caption{Empirical size and power of $R$ when data are generated from the Cox model.}
\label{wv_robustness_table}
\begin{tabular}{@{\extracolsep{\fill}} c c }
\hline
  \multicolumn{2}{c}{Empirical Size (Power)} \\ \hline
  \multicolumn{1}{c}{p=3, n=400} & \multicolumn{1}{c}{p=3, n=500} \\ \hline
 \multicolumn{1}{c}{0.049 (0.340)} & \multicolumn{1}{c}{0.049 (0.427)} \\ \hline
  \multicolumn{1}{c}{p=5, n=400} & \multicolumn{1}{c}{p=5, n=500} \\ \hline
\multicolumn{1}{c}{0.046 (0.565)} & \multicolumn{1}{c}{0.047 (0.665)} \\
\hline
\end{tabular}
\end{table}

\subsection{Empirical sizes and powers of $R$ and coxKM with a large number of SNPs sampled from the 1000 Genomes dataset}\label{sec_coxKM}

\begin{figure}[H]
\captionsetup[subfigure]{justification=centering}
\centering
\subcaptionbox{$R$}{\includegraphics[width=0.5\textwidth]{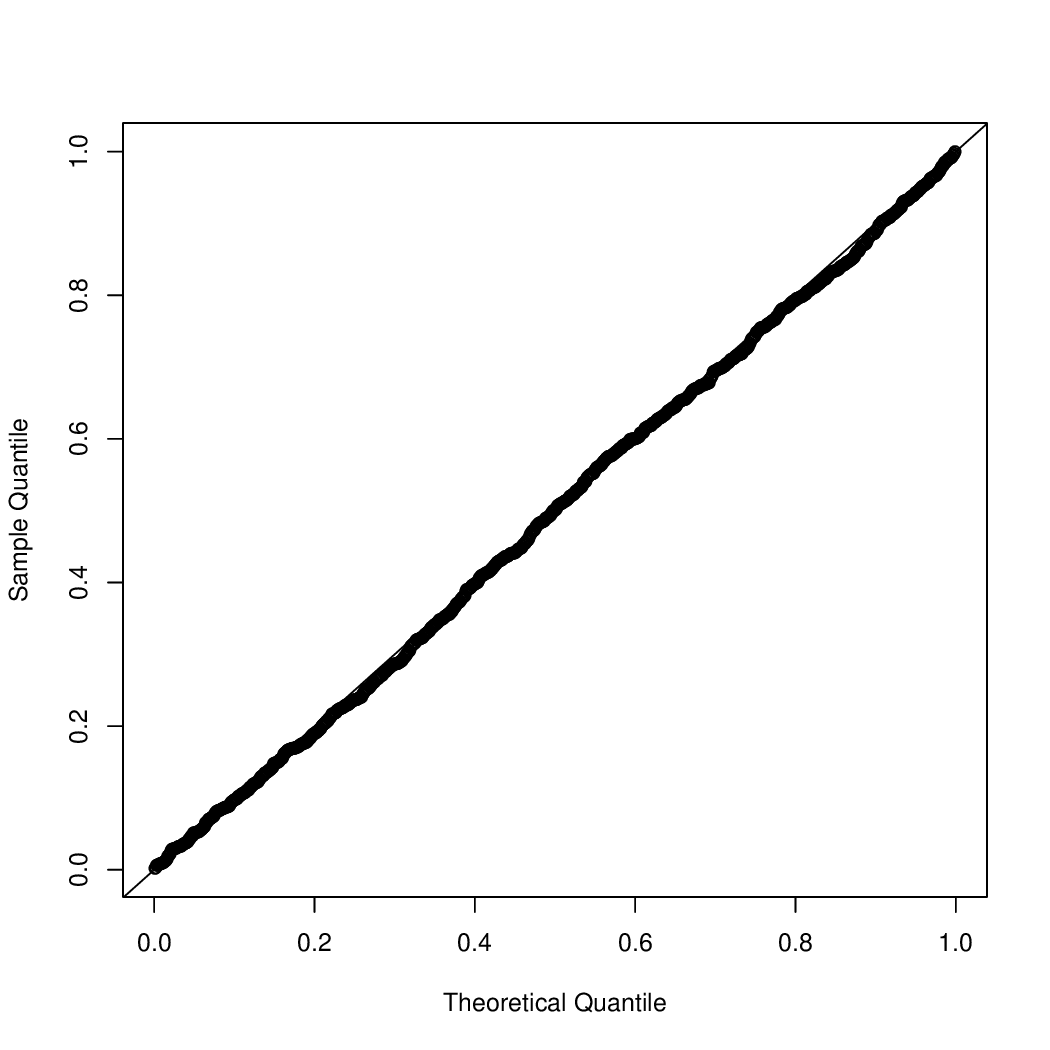}}%
\hfill 
\subcaptionbox{coxKM}{\includegraphics[width=0.5\textwidth]{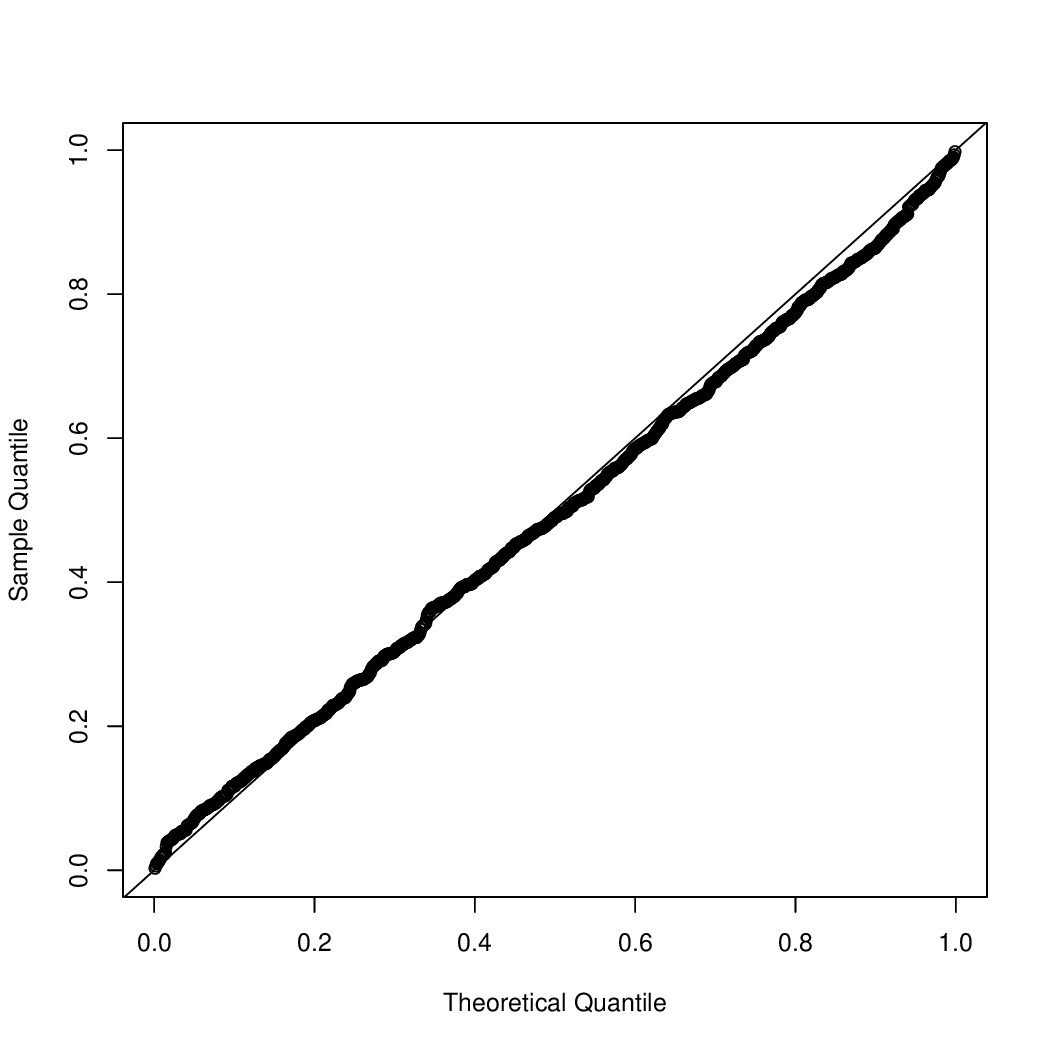}}%
\caption{Uniform Q-Q plots of the null p-values of $R$ and coxKM in testing genetic association when the survival data are generated from the AFT model, the size of the SNP set is large ($p=30$ and $n=400$), and the SNP genotypes are generated by sampling from the 1000 Genomes dataset.}
\label{wv_1kg_figure}
\end{figure}

In this simulation, we evaluated the empirical sizes and powers of our test $R$ and the Cox model-based test coxKM \citep{CaToLi} in testing genetic association when the size of the SNP set is large and the SNP genotypes were generated by sampling from the 1000 Genomes dataset (phase 3) \citep{1000G}. Except the size of the SNP set, the simulation setting is the same as for the simulation of association testing under no genetic heterogeneity. So the survival data were still generated from the AFT models. Table \ref{wv_1kg_table} shows that the empirical size of $R$ is close to the nominal level even when the SNP set size is $30$ and $R$ is a bit more powerful than coxKM. In contrast, coxKM is conservative when applied to survival data from AFT models when $n$ is relatively small considering $p$. Figure \ref{wv_1kg_figure} shows that $R$'s p-value follows the $U[0,1]$ distribution under the null, whereas the null distribution of coxKM's p-value is a bit deviated from the uniform distribution.

\begin{table}[H]
\setlength\tabcolsep{2pt}
\caption{Empirical sizes and powers of $R$ and coxKM with a large SNP set  sampled from the 1000 Genomes dataset.}
\label{wv_1kg_table}
\begin{tabular*}{\textwidth}{@{\extracolsep{\fill}} l *{3}{d{2.4}} }
\hline
 & \multicolumn{2}{c}{Empirical Size (Power)} \\ \cline{2-3}
 & \multicolumn{1}{c}{p=20, n=400} & \multicolumn{1}{c}{p=20, n=500} \\ \cline{2-3}
\multicolumn{1}{l}{$R$} & \multicolumn{1}{c}{0.049 (0.626)} & \multicolumn{1}{c}{0.049 (0.735)} \\
\multicolumn{1}{l}{coxKM} & \multicolumn{1}{c}{0.035 (0.614)} & \multicolumn{1}{c}{0.046 (0.705)} \\ \cline{2-3}
 & \multicolumn{1}{c}{p=30, n=400} & \multicolumn{1}{c}{p=30, n=500} \\ \cline{2-3}
\multicolumn{1}{l}{$R$} & \multicolumn{1}{c}{0.053 (0.830)} & \multicolumn{1}{c}{0.044 (0.886)} \\
\multicolumn{1}{l}{coxKM} & \multicolumn{1}{c}{0.030 (0.783)} & \multicolumn{1}{c}{0.042 (0.878)} \\
\hline
\end{tabular*}
\end{table}

\subsection{Runtimes of the proposed tests}\label{sec_runtimes}
We measured the runtimes of the proposed tests. The runtime measurements of the association tests $R$, $R_{\mbox{het}}$, $R^c$ and $R^c_{\mbox{het}}$ were taken in the settings of Table \ref{wv_1kg_table}. All the tests were executed on a workstation with a 28-core CPU at 2.40 GHz and 115 GB RAM. The programming language was R, but we used the R package Rcpp, which offers a seamless integration of R and C++, to reduce the runtime. Table \ref{runtime_table1}  shows the average runtimes of those tests. An interesting observation is that the number of genetic markers does not affect the runtimes of the association tests. This observation suggests that fitting the null model is the most time-consuming part of all the proposed tests.

\begin{table}[H]
\setlength\tabcolsep{2pt}
\caption{Runtimes of the proposed association tests in testing genetic effects in the absence of genetic heterogeneity and confounding.}
\label{runtime_table1}
\begin{tabular*}{\textwidth}{@{\extracolsep{\fill}} l *{5}{d{2.4}} }
\hline
 & \multicolumn{4}{c}{Average Runtime of Ten Replicates (seconds)} \\ \cline{2-5}
 & \multicolumn{1}{c}{$R$} & \multicolumn{1}{c}{$R_{\mbox{het}}$} & \multicolumn{1}{c}{$R^c$} & \multicolumn{1}{c}{$R^c_{\mbox{het}}$} \\
\hline
\multicolumn{1}{c}{p=20, n=400} & 4.28 & 4.80 & 5.25 & 5.50 \\
\multicolumn{1}{c}{p=30, n=400} & 4.92 & 4.70 & 5.60 & 5.52 \\
\hline
\multicolumn{1}{c}{p=20, n=500} & 6.60 & 7.05 & 8.14 & 8.31 \\
\multicolumn{1}{c}{p=30, n=500} & 6.62 & 6.47 & 7.95 & 8.26 \\
\hline
\end{tabular*}
\end{table}

\section*{Software}
The R codes that implement the proposed methods in this article are available at \url{https://github.com/didiwu345/Multi_Marker_AFT/}.

\section*{Funding}
This work was supported in part by the National Institutes of Health [R01DA043501,  R01LM012848, R56AG075803].

\section*{Acknowledgements}
The data for the application presented in this work were from the Religious Orders Study and the Rush
Memory and Aging Project supported by the National Institute of Aging
(U01AG61356, P30AG10161, R01AG15819, R01AG17917, R01AG019085 and R01AG30146). We are grateful to the Principal Investigator, David A. Bennett,
MD, for the authorization to use the data. The GWAS dataset (NG00029-ROSMAP GWAS) for the application was prepared, archived, and distributed by the National Institute on Aging Genetics of Alzheimer’s Disease Data Storage Site (NIAGADS) at the University of Pennsylvania (U24AG041689), funded by the National Institute on Aging. We also thank Professor Sy Han Chiou from Southern Methodist University for providing us the R code that implements the rank-based estimation for semiparametric accelerated failure time models under left truncation \citep{ChXu}.

\section*{Data Availability Statement}
The genotype data that support the findings of this study are available from
the National Institute on Aging Genetics of Alzheimer’s Disease Data Storage Site. Restrictions apply to the availability of these data, which were used under license in this paper. Data are available from \url{https://www.niagads.org/datasets/ng00029}  with the permission of the National Institute on Aging Genetics of Alzheimer’s Disease Data Storage Site. The Alzheimer's disease diagnosis data that support the findings in this paper are available from
 Rush Alzheimer's Disease Center. Restrictions apply to the availability of these data, which were used under license in this paper. Data can be requested at \url{www.radc.rush.edu}.

\section*{Conflict of Interest}
The authors declare that they have no conflict of interest.

\vspace*{-8pt}

\bibliographystyle{spbasic}
\bibliography{aft_references}

\end{document}